\documentclass[ip,twocolumn]{jpsj3}
\usepackage{txfonts}
\usepackage{color}
\usepackage{bm}

\newcommand{\tabref}[1]{Table.~\ref{#1}}
\newcommand{\equref}[1]{eq.~(\ref{#1})}
\newcommand{\figref}[1]{Fig.~\ref{#1}}
\newcommand{\secref}[1]{Sec.~\ref{#1}}

\title{
Deciphering long-range order in active matter:\\ Insights from swimming bacteria in quasi-2D and electrokinetic Janus particles
}

\author{Daiki Nishiguchi$^1$\thanks{nishiguchi@noneq.phys.s.u-tokyo.ac.jp}}
\inst{$^1$Department of Physics, The University of Tokyo, 7-3-1 Hongo, Bunkyo-ku, Tokyo 113-0033, Japan} 

\abst{
Emergent order resulting from spontaneous symmetry breakings has been a central topic in statistical physics. Active matter systems composed of nonequilibrium elements exhibit a diverse range of fascinating phenomena beyond equilibrium physics. One striking example is the emergent long-range orientational order in two dimensions, which is prohibited in equilibrium systems. The existence of long-range order in active matter systems was predicted first by a numerical model and proven analytically by dynamic renormalization group analysis. Experimental evidence for long-range order with giant number fluctuations has been provided in some experimental systems including microswimmers such as swimming bacteria and electrokinetic Janus particles. In this review, we provide a pedagogical introduction to the theoretical descriptions of long-range order in collective motion of active matter systems and an overview of the experimental efforts in the two prototypical microswimmer experimental systems. We also offer critical assessments of how and when such long-range order can be achieved in experimental systems. By comparing numerical, theoretical, and experimental results, we discuss the future challenges in active matter physics.}

\begin{document}
\maketitle

\section{Introduction}
\label{sec:Introduction}
Collections of self-propelled elements prevail in nature such as flocks of birds, schools of fish, cell populations, bacterial colonies, and cytoskeletal systems \cite{vicsek2012collective}. A framework of nonequilibrium statistical physics that tries to unveil the underlying universal laws in these novel nonequilibrium materials is now called active matter physics \cite{vicsek2012collective,marchetti2013hydrodynamics,chate2020dry,chate2022dry}. More specifically, active matter refers to nonequilibrium matter composed of elements that individually convert some sort of free energy into motion.

Active matter presents a rich variety of fundamental questions in statistical physics. Active matter systems are driven out of equilibrium at the individual element level and as such, they are far from equilibrium compared with other nonequilibrium systems that physics has traditionally dealt with.
Examples of traditional nonequilibrium systems include electrical conduction, thermal conduction, and fluid flow, in which nonequilibrium states are realized by imposing gradients of electric potential, temperature, or pressure as boundary conditions, respectively.
Other examples are glassy or granular systems, whose extremely slow dynamics prevent them from reaching thermal equilibrium within a realistic timescale. In contrast to these nonequilibrium states realized by boundaries or by their slow dynamics, active matter is intrinsically nonequilibrium with energy input and dissipation at the scale of each individual component, making it one of the most challenging classes of nonequilibrium systems. As such, active matter studies have revealed a diverse range of intriguing phenomena such as active turbulence at low Reynolds numbers \cite{alert2022active}, motility-induced phase separations (MIPS) in active Brownian particles \cite{cates2015motilityinduced}, and anomalous rheological responses including viscosity reduction up to negative values \cite{ramaswamy2010mechanics,lopez2015turning, chui2021rheology} and odd viscosity/elasticity \cite{fruchart2023odd}.

Among those properties stemming from activity, the emergence of long-range orientational order in two dimensions (2D) has been regarded as one of the hallmarks of collective motion in active matter.
This property, explored numerically and theoretically \cite{vicsek1995novel,toner1995longrange,toner1998flocks,ginelli2010largescale,mahault2019quantitative,chate2020dry,mahault2021longrange,chate2022dry}, has recently been evidenced by quantitative experiments \cite{nishiguchi2017longrange, tanida2020gliding, iwasawa2021algebraic}. Long-range order is associated with spontaneous symmetry breaking. For 2D systems in thermal equilibrium with short-range interactions, spontaneous symmetry breaking and subsequent long-range order are prohibited, which is now summarized as the Mermin-Wagner theorem \cite{mermin1966absence, hohenberg1967existence, mcbryan1977decay, frohlich1981absence}. The nonequilibrium nature of active matter, however, exempts it from the framework of this theorem, allowing the emergent long-range order in certain models and experimental settings, while the necessary conditions for long-range order to arise in 2D systems remain elusive and represent a fundamental question in the field of statistical physics \cite{tasaki2020hohenbergmerminwagnertype}.
The presence of long-range order, and associated anomalous giant fluctuations therein, was considered a universal property of collective motion due to the theoretical construction based solely on the symmetry of the system \cite{toner1995longrange, toner1998flocks, toner2005hydrodynamics, marchetti2013hydrodynamics, chate2020dry,chate2022dry}. However, experimental evidence of long-range order remained absent for more than two decades since the beginning of active matter physics \cite{vicsek1995novel,toner1995longrange}. Recent advances in active matter experiments are filling the gaps between theoretical predictions and experimental observations, and thus active matter physics is entering a new era with complementary advancements in theories and experiments.

In the experimental exploration of universal laws in active matter, microswimmers have been an invaluable experimental tool \cite{bechinger2016active,bar2019selfpropelled,aranson2022bacterial,alert2022active}.
A diverse range of systems has been employed to investigate the statistical properties of active matter, including shaken granular materials \cite{narayan2007longlived, deseigne2010collective, weber2013longrange, kumar2014flocking, soni2020phases}, flocking birds \cite{cavagna2010scalefree, attanasi2014information}, reconstituted cytoskeletons \cite{schaller2010polar, sanchez2012spontaneous, schaller2013topological, suzuki2015polar}, and cell migrations \cite{kawaguchi2017topological, saw2017topological, blanch-mercader2017turbulent}. However, microswimmers offer advantages in terms of controllability, reproducibility, experimental timescales, and the large number of elements in a system. In this review, among many microswimmer systems, we focus on two experimental systems, swimming bacteria and electrokinetic Janus particles, where the presence of true long-range orientational order has been quantitatively verified through finite-size scaling analysis.

This review is organized as follows. First, in \secref{sec:Microswimmers}, we introduce the basic behavior of swimming bacteria and electrokinetic Janus particles as useful tools in active matter physics. Then, in \secref{sec:LRO}, we review the theoretical results from the standard models of collective motion and discuss the experimentally observed long-range ordered phases from these theoretical perspectives. Finally, in \secref{sec:Conclusion}, we summarize the significance of experimental findings and discuss the remaining challenges in understandings of long-range order in active matter systems.

\section{Microswimmers as a tool for experimental exploration of active matter physics}
\label{sec:Microswimmers}
\subsection{Collective motion of swimming bacteria}
\label{sec:BacteriaDimension}

\begin{figure*}[tb]
\begin{center}
\includegraphics[width=0.8\textwidth]{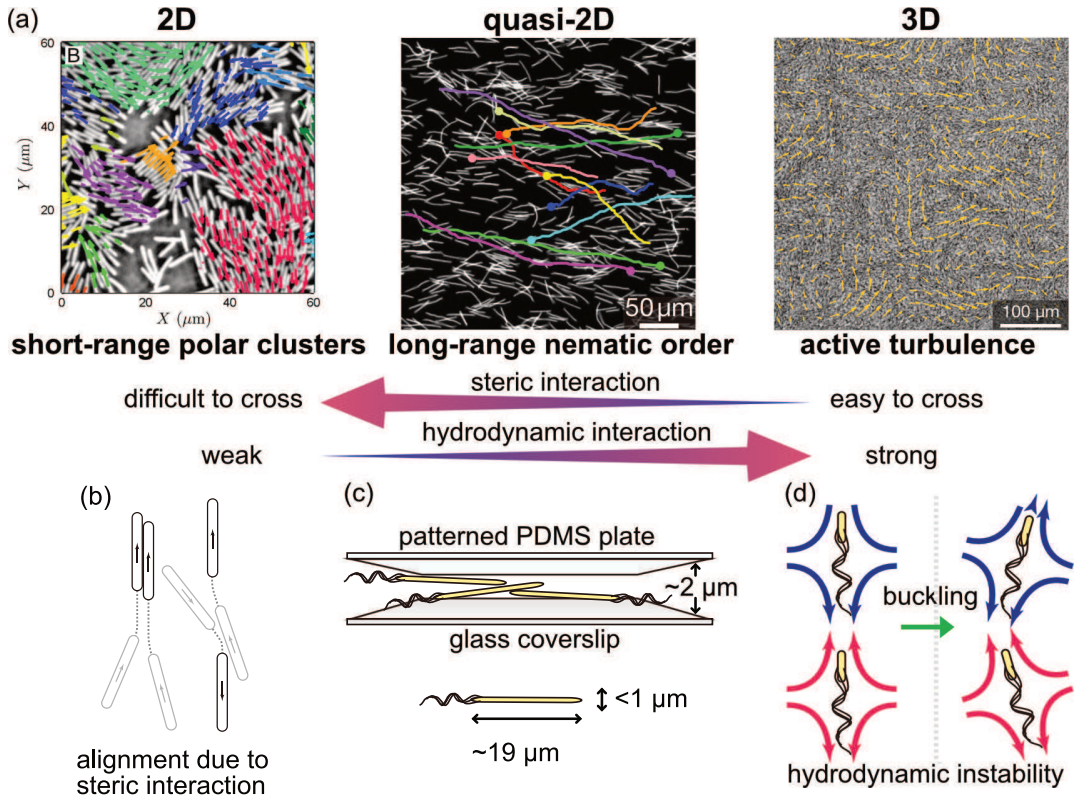}
\caption{Bacterial collective motion in different spatial dimensions. (a) Experimental snapshots and strengths of interactions. Vectors overlaid on 2D and 3D figures represent velocity fields. In the quasi-2D figure, trajectories of some selected bacteria are overlaid to visualize the nematic order. The figure for 2D is produced from Ref.~\citen{zhang2010collective}. The figure for quasi-2D is reproduced from Ref.~\citen{nishiguchi2017longrange}.
(b) Schematics of aligning steric interactions between rod-shaped bacteria. They work more strongly for smaller dimensions in which crossing each other becomes difficult. (c) Schematic of the quasi-2D experimental setup and typical length scales of elongated bacteria.
(d) Destabilizing hydrodynamic interactions become stronger for higher dimensions. As a result of the competition of the two interactions, different collective motion emerges depending on the spatial dimensions.
}
\label{fig:DimensionBacteria}
\end{center}
\end{figure*}

Among diverse mechanisms of bacterial motility \cite{miyata2020tree}, including swimming \cite{berg1973bacteria, lauga2009hydrodynamics, LaugaBook}, gliding \cite{peruani2012collective,nakane2013helical,copenhagen2021topological}, and twitching \cite{peruani2012collective, bonazzi2018intermittent, kennouche2019deep}, swimming with helical flagella is common to model bacteria such as {\it Escherichia coli} and {\it Bacillus subtilis} and therefore has been analyzed in detail using microhydrodynamics \cite{LaugaBook}. 
As such, these model swimming bacteria have been extensively employed as model active matter systems for studying collective motion \cite{bar2019selfpropelled, aranson2022bacterial}.
Their collective motion emerges through the competition between aligning steric interactions and destabilizing forces such as hydrodynamic interactions (\figref{fig:DimensionBacteria}(d)).
The rod-shaped bodies of these typical bacteria facilitate nematic interactions during collisions, inducing parallel or anti-parallel alignment akin to nematic liquid crystals composed of elongated molecules.
On the other hand, destabilization of this alignment is caused by extensile forces along the longitudinal direction that arises either by mechanical contact in the longitudinal directions \cite{wensink2012mesoscale,wensink2012emergent} or by dipolar hydodynamic flow \cite{dunkel2013minimal, reinken2018derivation}. These bacteria, referred to as pusher-type microswimmers \cite{LaugaBook}, swim by pushing the fluid behind them with their rotary helical flagella and in front of them with their bodies, creating dipolar flow around each bacterium. This pusher-type flow destabilizes straight configurations of bacterial alignment. The strengths of these steric and hydrodynamic interactions depend on the experimental realizations. Notably, the spatial dimensions of the systems strongly influence the emergent macroscopic collective behavior of swimming bacteria.

In the case of two-dimensional (2D) bacterial collective motion on a soft agar plate \cite{zhang2010collective}, the presence of the no-slip agar surface and the liquid-air interface stiffened by surfactant secreted by bacteria such as {\it B. subtilis} \cite{aranson2007model, lopez2014dynamics, peled2021heterogeneous} damps the hydrodynamic flow, allowing the steric interactions to dominate the dynamics. Consequently, the aligning steric interactions facillitate orientational order, forming locally aligned clusters. However, in such strictly 2D setups, bacteria are unable to cross each other \cite{ghosh2022cross}, which leads to the breakups of the aligned clusters after collisions. This process prohibits the growth of the aligned region, resulting in short-ranged polar clusters without long-range orientational order.

For dense bacterial suspensions in three dimensions (3D), spatio-temporally chaotic collective motion appears \cite{dombrowski2004selfconcentration}.
Unlike in 2D, the additional dimension allows aligned bacterial clusters to cross and avoid post-collision breakups.
Nevertheless, in 3D, the destabilizing hydrodynamic interactions become relevant due to the absence of boundaries to damp the hydrodynamic flow. This leads to continuous buckling and deformations of the bacterial director field, resulting in the chaotic state known as active turbulence \cite{alert2022active}.

Then, what happens in an intermediate case between 2D and 3D? 
Bacteria confined within a quasi-2D thin fluid layer between two plates turned out to exhibit long-range orientational order \cite{nishiguchi2017longrange}.
In this experiment, a highly quasi-2D environment with a gap width of $\sim 2\;\mathrm{\mu m}$, comparable to bacterial diameters and sufficiently shorter than bacterial body lengths, was realized so that bacteria interacted continuously with both of the plates.
Achieving a high ratio of the bacterial body lengths to the gap enables bacteria to swim straight by compensating their circular swimming close to a single plane wall \cite{lauga2006swimming, swiecicki2013swimming}. To facilitate this, the bacteria were artificially elongated by inhibiting cell divisions using an antibiotic, instead of reducing and fine-tuning the gap width at sub-$\mathrm{\mu m}$ precisions.
This unique experimental condition allowed bacteria to cross each other, which is not possible in strictly 2D systems. Consequently, as quantitatively demonstrated by varying the gap width of Hele-Shaw chambers \cite{ghosh2022cross}, qualitatively different bacterial alignment during collisions give rise to different macroscopic phases: short-range polar clusters for 2D and long-range nematic order for quasi-2D.
while the presence of two walls suppressed hydrodynamic interactions, resulting in the observed long-range nematic orientational order. This state can be discussed in the context of the Vicsek-style self-propelled rod model, a model extended from the original Vicsek model to represent polar particles with nematic interactions.
We will discuss in detail the implications of this ordered state in \secref{sec:EcoliLRO}.

\subsection{Electrokinetic Janus particles}
\label{sec:Janus}

\begin{figure*}[tb]
\includegraphics[width=\textwidth]{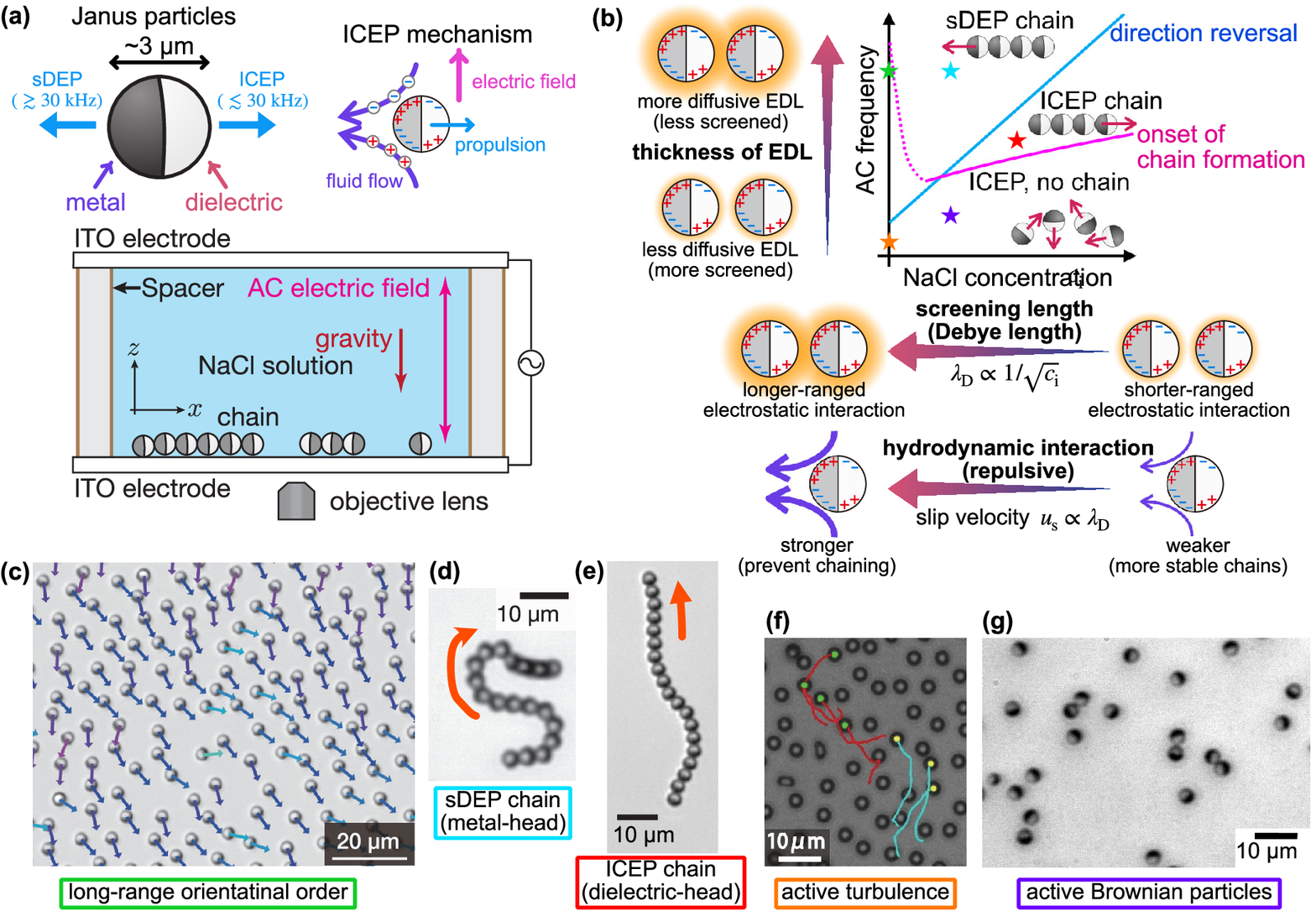}
\caption{Experiments of electrokinetic Janus particles and the phase diagram. (a) Schematics of Janus particles, ICEP mechanism, and the experimental setup. Cyan arrows represent the swimming directions of Janus particles. In the schematic of the ICEP mechanism, a moment of the electric field in the $+z$-direction is depicted. Fluid charged with counter ions is driven by the electric field, resulting in the propulsion of the particle in the opposite direction.
Figures reproduced and modified from Ref.~\citen{nishiguchi2018flagellar}.
(b) Schematic phase diagram of electrokinetic Janus particles. Cyan and magenta lines represent experimentally measured transition lines for the directional reversal and the onset of chain formation. When an AC electric field is applied in the $z$-direction, quadrupolar electric charge distribution is induced on the particles.
The ion concentration in the solution controls the thickness (Debye length) of the electric double layer (EDL) formed by counter-ions around the particles, which is represented by orange layers in the schematics. 
This also results in the change of hydrodynamic flow around the particles. For low enough ion concentrations, strong hydrodynamic flow prevents chain formation, which is represented by the dashed magenta line. Note that this magenta line is dashed in the low ion concentration regime to highlight the difficulty in experimentally determining the precise positions. This difficulty arises due to relatively large drifts of ion concentration, which result from the dissolution of ions from the experimental apparatus and/or from the air in contact with the suspension.
The frequency of the AC electric field also controls the thickness of EDL because the counter-ions need charging times to form EDL. Colored stars in the phase diagram correspond to the experimental snapshots in (c) to (g). (c) Long-range ordered phase observed at 1~MHz in pure water. Figure courtesy of Junichiro Iwasawa\cite{iwasawa2021algebraic}. (d) A chain moving toward the metal side with its head fixed to the substrate, showing flagellar beating behavior. Observed at 1~MHz in 0.1~mM NaCl solution. Reproduced and modified from Ref.~\citen{nishiguchi2018flagellar}. (e) Chains moving toward the dielectric side, observed at 100~kHz in 0.8~mM NaCl solution. Reproduced and modified from Ref.~\citen{nishiguchi2018flagellar}. (f) Turbulent collective motion with local order, observed at 1~kHz in pure water, with coating the surface of the electrode with a surfactant. Trajectories of some particles are overlaid to highlight the local order. Reproduced from \citen{nishiguchi2015mesoscopic}. (g) ABP-like behavior at low density in a weakly interacting regime, observed at 5~kHz in 0.1~mM NaCl solution. Reproduced from Ref.~\citen{poncet2021pair}.
}
\label{fig:JanusPhaseDiagram}
\end{figure*}

Active colloids represent another relevant experimental system in active matter.
Spherical colloidal particles whose two hemispheres have distinct physical properties are referred to as Janus particles \cite{bechinger2016active}  (\figref{fig:JanusPhaseDiagram}(a)), named after the two-faced Roman god Janus. 
The propulsion mechanisms of Janus particles are diverse but based on some sort of phoretic mechanisms arising from nonequilibriums transport. Such examples include self-diffusiophoresis by autocatalysis in hydrogen peroxide solutions \cite{paxton2004catalytic,howse2007selfmotile}, self-thermophoresis through a self-generated temperature gradient under laser irradiation \cite{jiang2010active}, and self-diffusiophoresis via the demixing of a critical binary mixture \cite{volpe2011microswimmers,buttinoni2012active}. In all of these mechanisms, the inherent polarities of Janus particles generate gradients or asymmetries in isotropic and homogeneous environments, leading to the propulsion of the particles.

Among many realizations of Janus particles, metallo-dielectric Janus particles under an AC electric field offer a highly controllable system that is useful for exploring the statistical physics of active matter systems \cite{suzuki2011validity, nishiguchi2015mesoscopic, nishiguchi2018flagellar, iwasawa2021algebraic, yan2016reconfiguring, boymelgreen2016propulsion, boymelgreen2022synthetic} (\figref{fig:JanusPhaseDiagram}). By varying the experimental parameters, the particle interactions as well as the direction of motion can be controlled, leading to a variety of macroscopic phases of active matter in a single experimental setup.

In this system, the Janus particles are suspended in either pure water or NaCl solution and subsequently sandwiched between two transparent indium-tin-oxide (ITO) electrodes. After waiting for a while so that the Janus particles sediment close to the bottom electrode, an AC electric field is applied vertically (in the $z$-direction). Consequently, the Janus particles start to swim in the horizontal plane (the $xy$-plane), perpendicular to the applied electric field. Importantly, the direction of motion is determined not by the external field but by the internal degree of freedom of each Janus particle, i.e. the metallo-dielectric polarity. Thus, the Janus particles move around by breaking the rotational symmetry around the $z$-axis of the external driving field. This feature distinguishes the active propulsion of the Janus particles from passive motion such as electrophoresis. The polarities of the Janus particles are kept perpendicular to the applied electric field due to hydrodynamic torque \cite{kilic2011inducedcharge}, enabling the realization of 2D dynamics of the Janus particles as they remain confined to the $xy$-plane.

The propulsion directions of Janus particles depend on the frequency of the applied AC electric field.
At low-frequencies $\lesssim$30~kHz, the Janus particles propel toward their dielectric side due to a mechanism called induced-charge electrophoresis (ICEP) (\figref{fig:JanusPhaseDiagram}(a)).
Under the electric field, surface charges are induced on a Janus particle, causing counter-ions in the suspending fluid to gather around the particle and form an electric double layer (EDL). The EDL screens the electric field, preventing it from penetrating the particle. This screened electric field has the tangential component along the surface of the particle and this drives the fluid within the EDL charged with the counter-ions. This flow is called the induced-charge electro-osmotic (ICEO) flow \cite{squires2004inducedcharge,squires2006breaking,bazant2010inducedcharge}. Because the induced charges are significantly larger on the metal side than on the dielectric side, the ICEO flow is much stronger on the metal side while the flow on the dielectric side is negligibly small (see \figref{fig:JanusPhaseDiagram}(a) right). This asymmetry in the hydrodynamic flow then propels the particle toward the dielectric side.
In contrast, at high-frequencies $\gtrsim$30~kHz, the particles move toward the metal side, first reported in Ref.~\citen{suzuki2011validity} and then also in other works \cite{yan2016reconfiguring, mano2017optimal, nishiguchi2018flagellar}.
This is a more subtle mechanism because the counter-ions do not have sufficient time to form an EDL within one period of the high-frequency field. 
Boymelgreen and coworkers have proposed a mechanism, self-dielectrophoresis (sDEP), where the Janus particle is driven by localized nonuniform electric field gradients induced between the particle and the bottom electrode \cite{boymelgreen2016propulsion, boymelgreen2022synthetic}.
In both ICEP and sDEP regimes, the self-propulsion speed is proportional to the squared amplitudes of the applied electric field \cite{gangwal2008inducedcharge,nishiguchi2018flagellar}.

The applied electric field induces quadrupolar charge distributions on the particles, viewed from the side, due to the different response times of the two hemispheres in both regimes \cite{nishiguchi2018flagellar}. Since the induced surface charges are screened by the EDL formed by the counter-ions in the suspending fluid, the electrostatic interactions between the particles work only when the inter-particle distances are smaller than the thickness of the EDL. When the EDL is sufficiently thin (cyan and red stars in \figref{fig:JanusPhaseDiagram}(b)), the quadrupole-quadrupole electrostatic interactions result in attractive forces between the metal and dielectric hemispheres of adjacent particles. This leads to the formation of chain structures, with tunable directions of motion depending on the frequency regimes \cite{nishiguchi2018flagellar} (\figref{fig:JanusPhaseDiagram}(d),(e)). These chains are capable of cargo transport by adhering to cargo through electrostatic interactions \cite{nishiguchi2018flagellar}. Intriguingly, the chain exhibits a beating behavior reminiscent of eukaryotic flagella found in sperms and algae when the head of a chain is fixed to the bottom electrode. This artificial flagellum was employed to verify theoretical  predictions \cite{sekimoto1995symmetry, camalet1999selforganized, chelakkot2013flagellar} for active filaments on the scaling relation between the beating frequency and the propulsion force of each component \cite{nishiguchi2018flagellar}.

By tuning the ion concentration of the solution and the frequency of the applied AC electric field, it is possible to control the thickness of the EDL and, consequently, observe various phases of collective motion other than the chain structures. In the low-frequency and low-ion concentrations regime (orange star in \figref{fig:JanusPhaseDiagram}(b)), the Janus particles exhibit polar alignment, which is destabilized by strong ICEO flow, resulting in active turbulence \cite{nishiguchi2015mesoscopic} (\figref{fig:JanusPhaseDiagram}(f)). The mechanism of this active turbulence is similar to bacterial turbulence as described in the previous section because the Janus particles create pusher-type flow fields at the lowest order \cite{nishiguchi2015mesoscopic}, as visualized experimentally \cite{peng2014inducedcharge}.
Increasing the ion concentration diminishes the electrostatic interactions and results in purely repulsive steric interactions dominating the dynamics (\figref{fig:JanusPhaseDiagram}(g), purple star in \figref{fig:JanusPhaseDiagram}(b)).
This weakly interacting state of the Janus particles was used to deepen the theoretical descriptions for the active Brownian particles (ABP) model \cite{callegari2019numerical}, one of the simplest models for active particles without alignment interactions. Experimental observation of the dilute regime with these parameters unraveled the existence of anomalous winged pair correlations with depletion wakes consisting of two depletion wings.
Analytical and numerical calculations characterized that the wings decay algebraically \cite{poncet2021pair}.
Around these parameters, Janus particles also form clusters at high density \cite{vanderlinden2019interrupted, zhang2021activeJanus}, which is comparable to the motility-induced phase separation of ABP \cite{cates2015motilityinduced}. Another collective phase was realized at the high-frequency and low-ion-concentration regime (green star in \figref{fig:JanusPhaseDiagram}(b)), in which a flock of Janus particles exhibited true long-range orientational order \cite{iwasawa2021algebraic} (\figref{fig:JanusPhaseDiagram}(c)). In this condition, the Janus particles are attracted to each other by their electrostatic interactions, while strong hydrodynamic flows around the particles prevent them from forming chains, giving rise to the flocking state.
This state can be interpreted in the framework of the Vicsek model, which will be discussed in \secref{sec:JanusLRO}.

From the experimental perspectives, the choice of the diameter of the Janus particles $2R\approx 3\;\mathrm{\mu m}$ is an essential factor to make the system suitable for statistical physics experiments. The relative strengths of the self-propulsion and the fluctuating noise, namely the P\'{e}clet number, determine the macroscopic behavior of the system \cite{poncet2021pair}. While the self-propulsion speed of the Janus particles scales proportionally to the diameter of the particle at least for the low-frequency ICEP regime \cite{squires2006breaking,gangwal2008inducedcharge}, the thermal fluctuations become weaker for the larger diameters. The estimates of the translational and rotational diffusion coefficient are given by the Einstein-Stokes relation as $D_t=\frac{k_\mathrm{B}T}{6\pi\eta R}$ and $D_r=\frac{k_\mathrm{B}T}{8\pi\eta R^3}$, respectively. Practically, the use of very small particles $2R\lesssim1\;\mathrm{\mu m}$ cause the Janus particles to escape from the two-dimensional layer close to the bottom electrode due to strong translational Brownian motion. This also results in chain and aggregate formation in the $z$-direction because of the induced surface charges of the particles, making the experimental observation difficult. For much larger diameters, since the rotational diffusion coefficient scales as $\propto R^{-3}$, the particles' behavior becomes strongly ballistic in a given experimental field of view and suppresses the stochasticity of the system, making the competition of order and fluctuations obscure.
The intermediate choice of $2R\approx 3\;\mathrm{\mu m}$ leads to $D_r\approx0.05\;\mathrm{s^{-1}}$ from the Einsten-Stokes relation, and, experimentally, larger values $D_r\approx 0.12\;\mathrm{s^{-1}}$ are reported \cite{poncet2021pair} due to e.g. the surface roughness of the electrodes or hydrodynamic torques introduced by other particles. This defines the characteristic rotational diffusion time scale $\sim$8 s. Considering that the typical swimming speed of the Janus particles used for experiments is about 5--10~$\mathrm{\mu m/s}$, the persistence length of the particles, 40--80 ~$\mathrm{\mu m}$, can be set sufficiently shorter than the linear size of a field of view. This enables us to distinguish and identify the macroscopic phases of collective motion in the system. For example, it allows us to discern whether the system possesses a global order far beyond the field of view or merely has short-range order.

The electrokinetic Janus particles are not only valuable for exploring statistical physics but also have a variety of practical applications.
For instance, their capability of suddenly switching the direction of motion by changing the frequency has been employed to steer the Janus particles toward a target by feedback controls \cite{mano2017optimal}. Another application is to use the Janus particles as microelectrodes for selective electroporation of bacteria \cite{wu2020active}. These particles have an additional ferromagnetic layer on the metal side for orientation control via an external magnetic field. One can manipulate the Janus particles to collect target bacteria on their surfaces, adjust the frequency of the AC electric field to halt the particles' motion while maintaining bacterial adhesion on the surfaces, and apply a continuous electric field to electroporate the adhered bacteria.
Recently, the electrokinetic Janus particles were combined with a unique substrate that undergoes a heat-induced amorphous-to-crystalline phase change \cite{nakayama2023tunable}. This phase change introduces memory to the system, leading to the emergence of memory-induced collective behavior. As a result, the Janus particles formed trails reminiscent of those of ants mediated by pheromone-based interactions.
These examples clearly demonstrate the potential of the electrokinetic Janus particles as a promising active matter system that can be further engineered for practical purposes.

\section{Long-range order from the viewpoint of the Vicsek-style models}
\label{sec:LRO}

\begin{figure}[tb]
\includegraphics[width=\columnwidth]{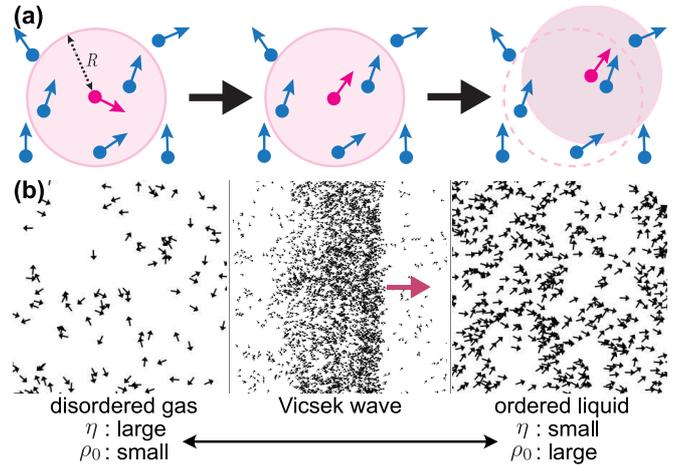}
\caption{(a) Schematic of the time evolution of a self-propelled particle in the Vicsek model. Each particle interacts with its neighbor with the radius $R$. (b) Phases of the Vicsek model at different densities $\rho_0$ and noise strengths $\eta$. The Visek model exhibits a discontinuous phase transition with a coexisting phase, in which a propagating ordered phase, called the Vicsek wave, is observed. The magenta arrow indicates the propagating direction of the Vicsek wave. Snapshots of the Vicsek model: Courtesy of Alexandre Solon.}
\label{fig:VicsekModel_PhaseDiagram}
\end{figure}

\subsection{The Vicsek model}
\subsubsection{Spirit of statistical physics and definition of the Vicsek model}
As discussed in previous sections, collective motion in active matter systems emerges through complicated interactions that vary significantly from one system to another. These interactions include forces and torques originating from steric, hydrodynamic, or electrostatic interactions in the cases of swimming bacteria and active colloids. For flocking animals, such as birds \cite{cavagna2010scalefree, attanasi2014information} and fish \cite{ito2022emergence,nishiguchi2022physics}, visual recognition and other sensory systems play important roles in their collective behavior. While an in-depth investigation of each system is certainly intriguing, statistical physics has sought universal properties common to different active matter systems. In analogy to the success in uncovering universality in ferromagnets by simple mathematical models such as the Ising model and the XY model, universal properties of collective motion, if any, should be manifest in simple models that capture the essential aspects of the system such as its symmetry.

In this spirit of statistical physics, Vicsek {\it et al.} devised a model in 1995 to extract the essence of the collective dynamics of self-propelled elements, which is now regarded as a standard model of collective motion \cite{vicsek1995novel,chate2020dry,chate2022dry}. In this model, a collection of spins (arrows) spontaneously moves in their own direction. Specifically, each component of the system is represented by a simple self-propelled particle that tries to align with its neighbors within the interaction radius (\figref{fig:VicsekModel_PhaseDiagram}(a)). In this original Vicsek model, the interaction has polar symmetry, meaning that it tends to align the particles in the same direction as in the spins in ferromagnetic models. We will discuss extensions to other interactions later in \secref{sec:VicsekStyleDifferentSymmetry}.
By denoting the direction of the $j$-th particle at time $t$ as $\theta_j^{t}$ and its position as $\boldsymbol{r}_j^{t}$, the dynamics of each particle is described as follows \cite{vicsek1995novel,ginelli2016physics,chate2020dry,chate2022dry}. For simplicity, we restrict ourselves to the 2D Vicsek model hereafter, but it is possible to extend the model to three or higher dimensions.
\begin{eqnarray}
\theta_j^{t+1}&=&\underbrace{\arg\sum_{k\sim j} \mathrm{e}^{\mathrm{i}\theta_k^{t}}}_{\text{average direction}} +\eta_j^t,\\
\boldsymbol{r}_j^{t+1}&=&\boldsymbol{r}_j^t+v_0 \boldsymbol{e}_{\theta_j^{t+1}},
\end{eqnarray}
where $\sum_{k\sim j}$ denotes the summation over neighboring particles, including the particle of the interest itself, within the interaction radius $R$, $\arg$ represents the argument of a complex number, $\eta_j^t$ is a white noise term uniformly distributed over the interval $[-\eta/2,+\eta/2]$ expressing that the particles do not align perfectly with the average direction of the neighbors, $\boldsymbol{e}_{\theta_j^{t}}$ is the unit vector in the direction of $\theta_j^{t}$, and $v_0$ is a constant representing the speed of each particle. Here, the unit of time is set as a single step of the dynamics since the Vicsek model is a discrete-time model. In the following, we will examine the results of numerical calculations for the time evolution of the Vicsek model in a $L\times L$ two-dimensional space with periodic boundary conditions and an average particle number density $\rho_0$.

\subsubsection{Flocking transition and long-range orientational order}
The Vicsek model contains four parameters: the self-propulsion speed $v_0$, the interaction radius $R$, the noise strength $\eta$, and the particle number density $\rho_0$. To investigate the behavior of the system, we need to choose control parameters. It is also desirable to reduce the number of parameters by physical inspections. For example, when the self-propulsion speed is as high as $v_0 > R$, the particles may not interact even if they come close to each other, which rationalizes the choice of $v_0 \leq R$. This inspection leads to normalizing the length scale to $R=1$ and setting $v_0 \simeq 1/2$, which do not affect the qualitative results \cite{chate2008collective}.
As a result, the remaining parameters are $\eta$ and $\rho_0$. The value of $\rho_0$ determines the number of particles interacting during a single step of the dynamics, corresponding to the effective strength of the interaction. The noise strength $\eta$ corresponds to the temperature in equlibrium models.

What does the macroscopic behavior of the system change when these parameters are varied? This is illustrated in \figref{fig:VicsekModel_PhaseDiagram}(b). When the noise $\eta$ is strong or the number density $\rho_0$ is low, the aligning interaction does not work sufficiently, resulting in a disordered phase where the particles move in uncorrelated random directions. On the other hand, by reducing the noise strength $\eta$ or increasing the particle number density $\rho_0$, the alignment interaction dominates the noise, resulting in an ordered phase where all the particles tend to move in the same direction on average. In active matter physics, this collective state with orientational order is called a flocking state, and the phase transitions to flocking states are referred to as flocking transitions.

Careful numerical calculations reveal a coexistence state in the parameter region between the disordered and ordered phases, where a band of a locally aligned flock propagates within the disordered phase \cite{chate2008collective,chate2020dry,chate2022dry}. This band is called the Vicsek wave. This coexistence of ordered and disordered phases signifies that the flocking transition in the Vicsek model is a discontinuous, first-order phase transition. Note that in the early numerical simulations of the Vicsek model, the system size was small, blurring the behavior at the transition point. Consequently, this phase coexistence was overlooked, and this phase transition was thought to be a continuous, second-order transition. Although we do not go into detail in this review, the flocking transition in the Vicsek model is currently considered to be best described as a nonequilibrium version of liquid-gas phase transitions \cite{chate2020dry,chate2022dry}. Unlike equilibrium liquid-gas phase separations where the system can surpass the phase boundary above the critical point without experiencing a phase transition, such a critical point in the Vicsek model goes to infinity as a result of different symmetries of the two phases. Based on this picture, the ordered and disordered phases are referred to as a polar liquid phase and a disordered gas phase, respectively. The Vicsek wave arises as a result of the liquid-gas phase separation.

\begin{figure}[tb]
\includegraphics[width=\columnwidth]{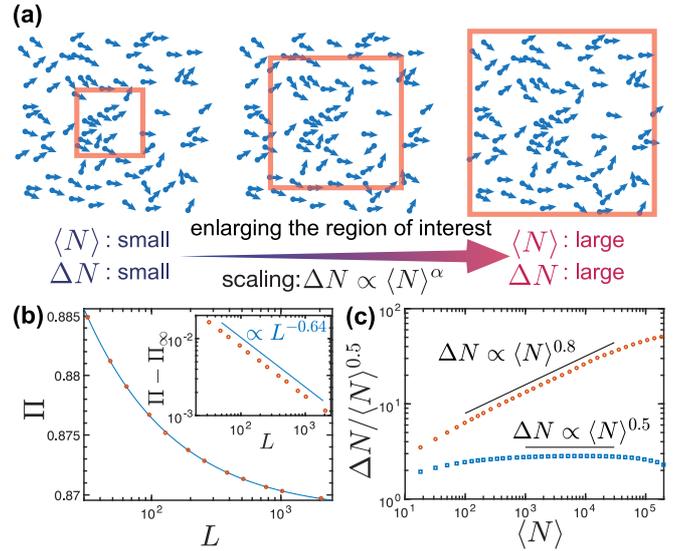}
\caption{(a) Schematic representation of the definition of number fluctuations and their exponent $\alpha$. Reproduced and modified from Ref.~\citen{nishiguchi2023natsugaku}. (b) Time-averaged values of polar order parameter $Pi$ as a function of the system size $L$ in the ordered phase of the Vicsek model, showing a decay slower than a power-law. (c) Number fluctuations calculated in the Vicsek model. In the ordered state (red circles), GNF was observed with the exponent $\alpha\simeq0.8$ (black), while normal fluctuations $\alpha=0.5$ were confirmed in the disordered state (blue squares). The system sizes are  $L=128$ for the ordered state and $L=512$ for the disordered state. Data used in (b) and (c) are provided by Beno\^it Mahault \cite{chate2020dry,chate2022dry}.}
\label{fig:VicsekModel_LROGNF}
\end{figure}

One of the most important properties of the ordered flocking phase in the Vicsek model is the emergence of true long-range orientational order \cite{toner1995longrange,chate2020dry}. The flocking transition in the Vicsek model is a phase transition from the disordered phase with continuous rotational symmetry to the flocking state where this symmetry is spontaneously broken. Notably, although individual particles in the Vicsek model interact only through short-range interactions within a radius $R$, a global order develops through the spontaneous breaking of rotational symmetry. The existence of long-range orientational order in the Vicsek model has been demonstrated both by the dynamic renormalization group analysis in a hydrodynamic description for the Vicsek model \cite{toner1995longrange} and by numerical calculations \cite{chate2020dry}. The numerical validation can be performed by examining the decay of a polar order parameter $\Pi=\langle | \langle \exp{(i\theta_j^t)\rangle_j | \rangle_t}$ as a function of the system size $L$ (see \figref{fig:VicsekModel_PhaseDiagram}(c)), where brackets $\langle \cdot\rangle_j$ and $\langle \cdot\rangle_t$ denote averages over all the particles and time respectively. This finite-size scaling analysis shows that the polar order $\Pi$ decays slower than a power law and converges to a positive finite value, signifying that the polar order persists even for the infinite system-size limit.

As a matter of fact, the discovery of this long-range orientational order was one of the driving forces behind the increased interest in active matter in the field of nonequilibrium statistical physics \cite{vicsek1995novel}.
In equilibrium systems, the emergence of long-range order resulting from spontaneous breakings of any continuous symmetry, such as continuous rotational symmetry, is prohibited in two-dimensional systems with short-range interactions. This is summarized as the Hohenberg-Mermin-Wagner theorem \cite{mermin1966absence, hohenberg1967existence, mcbryan1977decay, frohlich1981absence}. Therefore, for example, in the equilibrium classical XY model, long-range orientational order is not realized, and instead, quasi-long-range order is observed, where the correlation functions decay algebraically.
As a non-equilibrium model, the Vicsek model is, of course, beyond the scope of the Hohenberg-Mermin-Wagner theorem. Nonetheless, the discovery of long-range orientational order in a two-dimensional system with short-range interactions was a big surprise for statistical physics. This leads to a new question: Under what conditions can a system escape the framework of the Hohenberg-Mermin-Wagner theorem and exhibit long-range order? This is a fundamental question in statistical physics that continues to intrigue researchers not only in active matter physics but also in other fields of physics.
Regarding nonequilibrium systems other than active matter, for example, long-range order has recently been reported in the $O(2)$ model advected by shear flow \cite{nakano2021longrange} and in the XY model but with nonreciprocal interactions \cite{loos2023longrange}.
Recent advances in understanding the underlying mechanisms of emergent long-range order suggest that the nonequilibrium nature of the system, i.e., the breaking of detailed balance, plays a crucial role \cite{dadhichi2020nonmutual, tasaki2020hohenbergmerminwagnertype}, but we have not reached a systematic understanding yet.

\subsubsection{Giant number fluctuations}
Another hallmark property of collective motion in active matter is giant number fluctuations (GNF) \cite{narayan2007longlived,ginelli2016physics,chate2020dry}. The long-range ordered phases showing GNF in the Vicsek model and related systems are sometimes referred to as the Toner-Tu-Ramaswamy phases.
Let us first define what number fluctuations are.
Suppose we choose an observation area within a system. We then monitor the time series of the number of particles $N(t)$ in this region and calculate the mean value $\langle N \rangle$ and the standard deviation (fluctuation) $\Delta N$.
Next, consider increasing the size of this observation area; both the mean value and the fluctuation will increase. Now we are posed with a question: What relationship exists between these two quantities? Typically, for large system-size limits, we observe asymptotic power-law behavior, $\Delta N \propto \langle N \rangle^{\alpha}$. Note that, here we defined the exponent $\alpha$ based on the scaling of the standard deviation, but in some literature the scaling of the variance is used, resulting in the difference of a factor of two.
In equilibrium systems or completely random systems, the central limit theorem assures the relationship $\Delta N \propto \langle N \rangle^{0.5}$.  
This fluctuation with $\alpha = 0.5$ is regerred to as normal number fluctuations.
Then, what about in the Vicsek model? In the disordered phase, with little correlations in the particle positions and almost random motion of the particles, the central limit theorem holds and normal fluctuations with $\alpha = 0.5$ are observed. However, in the long-range ordered phase above the flocking transition points, the situation is more interesting. While the density distribution of the long-range ordered phase also appears to be uniform at first glance, the presence of long-range correlations makes the system deviate from the central limit theorem and exhibit fluctuations with a significantly larger exponent, $\alpha \simeq 0.8$. This is called giant number fluctuations. Note that, as will be discussed later, in the Toner-Tu calculations in 1995\cite{toner1995longrange}, it was shown by dynamic renormalization group analysis that the value is exactly 0.8 in two dimensions, but later reanalysis and large-scale numerical calculations have shown that this value is not accurate \cite{toner2012reanalysis,mahault2019quantitative}.

The fascinating aspect of GNF observed in the Vicsek model resides in the fact that it originates from the spontaneous breaking of rotational symmetry. In systems where rotational symmetry is spontaneously broken, such as the long-range ordered phase of the Vicsek model, there is no restoring force acting against perturbations that simultaneously rotate all the particles' directions uniformly because the relative angles between the particles do not change, keeping the interaction energy constant. As a result, very slow, long-wavelength fluctuations arise in the particle orientation field. This is a kind of the Nambu-Goldstone modes, which causes power-law correlations to appear in quantities such as particle density fields and orientation fields \cite{forster1990hydrodynamic,ginelli2016physics}. 
In the Vicsek model, the density field and orientation field, i.e. velocity field, are coupled, so we can intuitively understand that GNF is observed when the fluctuations of the orientation field are added to the fluctuations of the density field in the next time step \cite{ginelli2016physics}. Therefore, GNF in the Vicsek model can be regarded as a reflection of the mathematical properties arising from the spontaneous breaking of the system's rotational symmetry. This GNF mechanism based on the Nambu-Goldstone mode can also be well understood by examining the active Ising model, a lattice model for collective motion where spins hop on the lattice based on their orientations \cite{solon2013revisiting,solon2015flocking, solon2015phase}. The active Ising model has only discrete rotational symmetry based on the lattice structures, and thus no such Nambu-Goldstone modes are present even in the ordered phase where the spins are aligned and move together. This results in the absence of GNF, leading to normal number fluctuations.

So far, although we have defined the average $\langle N \rangle$ and fluctuations $\Delta N$ based on time-averaging, we can also define them based on spatial averaging at a single time point as far as the system is ergodic. From an experimental perspective, however, spatial inhomogeneities sometimes cannot be completely removed from the experimental setup. In such cases, time-averaging may provide more accurate estimates with fewer errors than spatial averaging. Conversely, when a system's behavior is unstable and unsteady, as with proliferating, deteriorating, or dying bacteria or cells, spatial averaging can be a better estimator.

Since GNF has been regarded as a hallmark of collective motion in active matter, it has been tested in numerical simulations and various experimental systems.
However, GNF needs to be carefully assessed and interpreted. For example, consider measuring the number fluctuations in the phase-coexistent state of the Vicsek model on the transition point. In this case, the number of particles fluctuates significantly depending on whether the band of the Vicsek wave passes through the observation area or not. This could result in seemingly giant fluctuations. In this way, the presence of spatially localized structures or clusters may generate giant fluctuations but not in the sense of the GNF observed in the Vicsek model.
Critical reviews of GNF observations reported in existing experiments can be found in the Supplemental Material of Ref.~\citen{nishiguchi2017longrange} and in Ref.~\citen{NishiguchiSpringerTheses}.

\subsubsection{Hydrodynamic description: the Toner-Tu theory}
The detailed analysis of the Vicsek model has been performed through the corresponding hydrodynamic description: the Toner-Tu equations \cite{toner1995longrange,toner1998flocks,toner2012reanalysis}. These partial differential equations for the coarse-grained velocity field $\boldsymbol{v}$ and particle number density field $\rho$ describe the macroscopic behavior of the Vicsek model. The whole theory, including predictions based on these equations, is called the Toner-Tu theory.
Toner \& Tu conceived these phenomenological equations by writing down the terms allowed by the symmetries of the system up to the lowest order in spatial and time derivatives, a similar approach used for deriving hydrodynamic equations for liquid crystals. Specifically,
\begin{align}
&\partial_t \bm{v}+\lambda_1 (\bm{v}\cdot \bm{\nabla})\bm{v}+\lambda_2(\bm{\nabla}\cdot\bm{v})\bm{v}+\lambda_3\bm{\nabla}(|\bm{v}|^2) \nonumber \\
&=a_0 \bm{v}-b_0 |\bm{v}|^2\bm{v}-\bm{\nabla}P_1-\bm{v}(\bm{v}\cdot\bm{\nabla}P_2) \nonumber \\
&+D_B\bm{\nabla}(\bm{\nabla}\cdot\bm{v})+D_T\nabla^2\bm{v}+D_2(\bm{v}\cdot\bm{\nabla})^2\bm{v}+\bm{f}, \label{eq:TonerTuEquation}\\
&\partial_t \rho+\bm{\nabla}\cdot(\bm{v}\rho)=0,\\
&P_1=\sum_{n=1}^\infty\sigma_n(|\bm{v}|)(\rho-\rho_0)^n, \label{eq:TonerTuPressure1}\\
&P_2=\sum_{n=1}^\infty\mu_n(|\bm{v}|)(\rho-\rho_0)^n, \label{eq:TonerTuPressure2}\\
&\langle f_i(\bm{r},t) f_j(\bm{r}',t')\rangle=\Delta \delta_{ij}\delta^d(\bm{r}-\bm{r}')\delta(t-t'). \label{eq:TonerTuNoise}
\end{align}
The coefficients $\lambda_1$, $\lambda_2$, $\lambda_3$, $a_0$, $b_0$, $P_1$, and $P_2$ in \equref{eq:TonerTuEquation} are, in general, functions of the local density field $\rho$ and the local magnitude of the velocity $|\bm{v}|$. Their corresponding Taylor expansions are explicitly given in eqs.~(\ref{eq:TonerTuPressure1}) and (\ref{eq:TonerTuPressure2}).
$P_1$ is an isotropic pressure term, which also exists in conventional fluids, and $P_2$ is an anisotropic pressure term, which is allowed in active matter systems. $b_0$, $D_B$, $D_T$, and $D_2$ are all positive, and the terms with $D_B$, $D_T$, and $D_2$ represent the diffusions of the velocity field due to particle interactions.
The term $a_0\bm{v}-b_0|\bm{v}|^2\bm{v}$ represents the self-propulsion, which can be derived from a Ginzburg-Landau type potential $-\frac{1}{2}a_0|\bm{v}|^2+\frac{1}{4}b_0|\bm{v}|^4$. This gives a constant velocity by spontaneously breaking the symmetry, describing the flocking transitions, as $a_0$ changes its sign: no spontaneous velocity for $a_0<0$ and spontaneous flow $v_0=\sqrt{\frac{a_0}{b_0}}$ for $a_0>0$. Lastly, $\bm{f}$ is uncorrelated white Gaussian noise, with its $i$-th component $f_i$ defined in \equref{eq:TonerTuNoise} and its variance $\Delta$, and $d$ is the dimension of space.

The terms involving $\lambda$ on the left side of \equref{eq:TonerTuEquation} represent advection. In the Navier-Stokes equation, Galilean invariance and the moment conservation need to hold, requiring $\lambda_1=1$ and $\lambda_2=\lambda_3=0$. However, in descriptions of active matter, Galilean invariance can be generally absent, and the total momentum is not conserved. 
For example, a school of fish swimming relative to the surrounding water may be considered to be in their ``absolute rest frame''. Therefore, if only the motion of the fish is taken into account by disregarding the motion of the water, the total momentum of the system is not conserved. The same situation applies to the air for a flock of birds and the ground for a herd of cattle, all of which act as momentum sinks when only looking at the dynamics of the self-propelled elements. In the Vicsek model, each particle moves at a constant speed $v_0$ relative to their absolute rest frame, and momentum is not a conserved quantity.

Analyzing the Toner-Tu equations with dynamic renormalization group technique has proven the existence of long-range order \cite{toner1995longrange,toner1998flocks}. In addition, predictions on power-law behaviors and their exponents of correlation functions and sound modes have also been obtained \cite{toner1995longrange,toner1998flocks,tu1998sound}. Since the Toner-Tu equations and their theoretical framework rely solely on the symmetries of the system, these predicted properties are expected to be universal, at least to some extent. 
However, Toner's reanalysis in 2012 \cite{toner2012reanalysis} reported several nonlinear terms overlooked during the original Toner \& Tu's 1995 derivation \cite{toner1995longrange, toner1998flocks} (hereinafter Toner-Tu 1995). Because of their undetermined relevance in the renormalization group analysis,
while the result of the existence of long-range order remains unchanged, the values of the various exponents derived in 1995 may be subject to change. There are indeed some discrepancies between these results and those of the recent largest-scale numerical simulations of the Vicsek model with billions of particles \cite{mahault2019quantitative} and experiments with electrokinetic Janus particles \cite{iwasawa2021algebraic}, as we will discuss in \secref{sec:JanusLRO}.

We note that the Toner-Tu equations have also been derived using the Boltzmann equation \cite{peshkov2012nonlinear, bertin2013mesoscopic, peshkov2014boltzmannginzburglandau,chate2020dry}. This approach elucidates how each coefficient in the Toner-Tu equations depends on the microscopic parameters of the Vicsek model. There, however, remain some difficulties such as the absence of some terms allowed by symmetry. This originates from the assumptions made during the derivation of the hydrodynamic equations through the Boltzmann equation: taking the dilute limit, assuming binary collisions, the molecular chaos assumption, and the truncations of higher-order terms. Nonetheless, most of the terms in the Toner-Tu equations have been successfully derived with this approach.

\subsubsection{Scaling relations and exponents}
\label{sec:VicsekScalingExponents}
In this subsection, we summarize the Toner-Tu 1995 predictions on various types of correlation functions and their exponents based on the dynamic renormalization group analysis \cite{toner1995longrange,toner1998flocks}.
As discussed previously, the long-range ordered phase above the flocking transition points exhibits power-law behavior in various correlation functions. These power-law scale-free correlations are important quantifiers of the system because they are associated with the Nambu-Goldstone mode arising from the spontaneous breaking of the rotational symmetry during the flocking transitions.
Even though the predictions made in Toner-Tu 1995 \cite{toner1995longrange,toner1998flocks} may not be perfect as discussed in the previous subsection \cite{toner2012reanalysis}, they provide a useful baseline for evaluating experimental and/or numerical results.

The dynamic renormalization group analysis for the ordered phase of the Toner-Tu equation (\equref{eq:TonerTuEquation}) seeks to identify the exponents that remain invariant under the following scale transformations \cite{toner1995longrange,toner1998flocks,toner2005hydrodynamics,toner2012reanalysis}: $\bm{x}_{\perp} \rightarrow b \bm{x}_{\perp}$, $x_{\|} \rightarrow b^{\zeta} x_{\|}$, $t \rightarrow b^{z} t$, $\bm{v}_{\perp} \rightarrow b^{\chi} \bm{v}_{\perp}$, $\delta \rho \rightarrow b^{\chi_{\rho}} \delta \rho$. Here, $\perp$ and $\|$ respectively denote the directions perpendicular and parallel to the global orientational order of the flock. Hence, $\bm{x}_{\perp}$ and $x_{\|}$ represent the coordinates perpendicular to and along the orientational order, respectively. Note that, in spatial dimensions $d>2$, $\bm{x}_{\perp}$ becomes a $(d-1)$-dimensional vector, and therefore the vector notation is used. To proceed with the calculations, we focus on fluctuations in the velocity field and the density field,
\begin{align}
&\rho(\bm{r}, t)=\rho_{0}+\delta \rho(\bm{r}, t), \\
&\bm{v}(\bm{r}, t)=\left(\langle v\rangle+\delta v_{\|}(\bm{r}, t)\right) \hat{\bm{e}}_{\|}+\delta \bm{v}_{\perp}(\bm{r}, t),
\end{align}
where $\hat{\bm{e}}_{\|}$ is the unit vector along the global orientational order of the flock, and $\delta \bm{v}_{\perp}$ represents fluctuations in the velocity field perpendicular to the global orientational order. Because it turns out that $\delta \rho$ scales as $\delta \rho \sim \delta \bm{v}_{\perp}$, a relation $\chi = \chi_{\rho}$ holds \cite{toner1995longrange,toner1998flocks}. Therefore, we only consider $\chi$ hereafter. The asymptotic behavior of their correlation functions is given by the following expressions, where $\delta\hat{\rho}(\bm{q},t)$ is the Fourier transform of $\delta\rho$, and $\bm{q}$ represents the wave vector:
\begin{align}
C_{v}(\bm{R}) &:=\left\langle\delta \bm{v}_{\perp}(\bm{r}+\bm{R}, t) \cdot \delta \bm{v}_{\perp}(\bm{r}, t)\right\rangle \nonumber \\ 
&\sim \begin{cases}\left|R_{\|}\right|^{2 \chi / \zeta} & \left( \left|\bm{R}_{\perp}\right|^{\zeta} \ll \left|R_{\|}\right|  \right) \\
\left|\bm{R}_{\perp}\right|^{2 \chi} & \left(\left|R_{\|}\right| \ll\left|\bm{R}_{\perp}\right|^{\zeta} \right)\end{cases}
\label{eq:TonerTuVelocityCorrelation}
\end{align}

\begin{align}
\hat{C}_{v}(\bm{q}) &:=\left\langle\delta |\bm{v}_{\perp}(\bm{q}, t)|^2\right\rangle \nonumber \\ 
&\sim \begin{cases}\left|q_{\perp}\right|^{1-d-\zeta-2\chi} & \left( \left|q_{\|}\right| \ll\left|\bm{q}_{\perp}\right|^{\zeta} \right) \\
\left|\bm{q}_{\|}\right|^{(1-d-\zeta-2\chi)/\zeta} & \left(\left|\bm{q}_{\perp}\right|^{\zeta} \ll\left|q_{\|}\right| \right)\end{cases}
\label{eq:TonerTuVelocityCorrelationFourier}
\end{align}

\begin{align}
&\hat{C}_{\rho}(\bm{q}) :=\left\langle|\delta \hat{\rho}(\bm{q}, t)|^{2}\right\rangle \nonumber \nonumber\\
& \sim \begin{cases}\left|\bm{q}_{\perp}\right|^{1-d-\zeta-2 \chi} & \left(\left|q_{\|}\right| \ll\left|\bm{q}_{\perp}\right|\right) \\
\left|q_{\|}\right|^{-2}\left|\bm{q}_{\perp}\right|^{3-d-\zeta-2 \chi} & \left(\left|\bm{q}_{\perp}\right| \ll\left|q_{\|}\right| \ll\left|\bm{q}_{\perp}\right|^{\zeta}\right) \\
\left|q_{\|}\right|^{-2+(1-d-\zeta-2 \chi) / \zeta}\left|\bm{q}_{\perp}\right|^{2} & \left(\left|\bm{q}_{\perp}\right|^{\zeta} \ll\left|q_{\|}\right|\right)\end{cases} \label{eq:TonerTuDensityCorrelation}
\end{align}
As can be seen in these expressions, the exponent $\zeta$ appears as a parameter characterizing the anisotropy of the decays of the correlation functions. Therefore, $\zeta$ is called an anisotropy parameter.

The exponents in the correlation functions have connections to other exponents.
An exponent $\beta:=-(1-d-\zeta-2\chi)$ is defined in the $\left|q_{\|}\right| \ll\left|\bm{q}_{\perp}\right|$ limit of eqs. \eqref{eq:TonerTuVelocityCorrelationFourier} and \eqref{eq:TonerTuDensityCorrelation}, connecting the GNF exponent $\alpha$ as $\alpha=1/2+\beta/2d$. This is because the density correlation function in Fourier space, i.e. the structure factor, is directly linked to density fluctuations \cite{ginelli2016physics}. Another relation arises from the superdiffusion. In the ordered phase of the Vicsek model, positions of individual particles $\bm{r}_i(t)$ exhibit superdiffusion in the direction perpendicular to the global orientational order \cite{toner1995longrange,toner1998flocks,tu1998sound,chate2008collective}. Quantifying their superdiffusive behavior via mean square displacement, $\left\langle\left[\bm{r}_{i, \perp}\left(t+t_{0}\right)-\bm{r}_{i, \perp}\left(t_{0}\right)\right]^{2}\right\rangle \sim t^{\nu}$, defines an exponent $\nu$, given by $\nu=\max (2+2 \chi / \zeta, 1)$. This means superdiffusion $\nu=4/3$ for $d=2$ and normal diffusion $\nu=1$ for $d=3$.

The dynamical exponent $z$ can be estimated from the dispersion relations of sound modes, i.e. the decays of fluctuations of the density or orientation fields that propagate within the flock \cite{toner1998flocks, tu1998sound,toner2012reanalysis}. Flocking states with true long-range order
can host sound modes despite their overdamped dynamics \cite{cavagna2015silent}.
To quantify the sound modes and the exponent $z$, we may calculate their space-time correlation functions in Fourier space, namely their power spectra, and compare them with theoretical results given by \cite{toner2012reanalysis, mahault2019quantitative},
\begin{align}
&\left\langle|\delta \rho(\bm{q}, \omega)|^2\right\rangle=\frac{\rho_0^2 \Delta}{\mathcal{S}(\bm{q}, \omega)} |\bm{q}_{\perp}|^2, \label{eq:SpaceTimeDensityCorr} \\
&\left\langle\left|\delta \mathbf{v}_{\perp}(\bm{q}, \omega)\right|^2\right\rangle=\frac{\left(\omega-v_2 q_{\|}\right)^2\Delta}{\mathcal{S}(\bm{q}, \omega)}+\frac{(d-2)\Delta}{\mathcal{S}_T(\bm{q}, \omega)},
\end{align}
with the denominators defined as,
\begin{align}
&\mathcal{S}(\bm{q},\omega)=\left\{\left[\omega-c_{+}(\theta_{\bm{q}}) q\right]^2+\varepsilon_{+}^2(\bm{q})\right\}\left\{\left[\omega-c_{-}(\theta_{\bm{q}}) q\right]^2+\varepsilon_{-}^2(\bm{q})\right\},\\ 
&\mathcal{S}_T(\bm{q}, \omega)=\left[\omega-c_T(\theta_{\bm{q}}) q\right]^2+\varepsilon_T^2(\bm{q}),
\end{align}
where $\theta_{\bm{q}}$ is the angle between the global order and $\bm{q}$, and $c_\pm(\theta_{\bm{q}})$ represents the propagating sound velocities in the $\theta_{\bm{q}}$ and $\theta_{\bm{q}}+\pi$ directions respectively. Detailed definitions of $v_2, \varepsilon_\pm, c_{\pm, T}(\theta_{\bm{q}})$ can be found in Ref.~\citen{toner2012reanalysis}. Importantly, these correlation functions have peaks at $\omega=c_{\pm}(\theta_{\bm{q}})$ with the widths of $\varepsilon_{\pm}$. The scaling relations between $\varepsilon_{\pm}$ and $\bm{q}$ represent the decay, or damping, of the sound modes during propagation, and the renormalization analysis gives,
\begin{align}
\varepsilon_{\pm} \sim 
 \begin{cases} |\bm{q}_\perp|^z & \left( |\bm{q}_\perp|^\zeta \gg q_\| \right) \\
q_\|^{z/\zeta} & \left( |\bm{q}_\perp|^\zeta \ll q_\| \right)\end{cases}.
\end{align}
Therefore, inspecting the peak width as a function of $\bm{q}$ provides an estimate of the dynamical exponent $z$.
While the exponent $z=2$ corresponds to the dispersion of the sound mode with normal diffusive behavior, in the flocking state such diffusion is suppressed, leading to less diffusive propagation with $z<2$. Recently, sound wave modes with $z<2$ have been detected in experiments with self-propelled colloids such as colloidal Quincke rollers under a DC electric field \cite{geyer2018sounds} and electrokinetic Janus particles \cite{iwasawa2021algebraic}.
Among the exponents introduced so far, only three of them, $\chi$, $\zeta$, and $z$, are independent, from which the remaining exponents $\beta$, $\alpha$, and $\nu$ can be calculated. In addition, a hyperscaling relation $z=\beta$ has been predicted and found to hold in numerical calculations \cite{mahault2019quantitative}.

\begin{table*}[tb]
{
 \begin{center}
 \caption{\label{tab:exponents}
The values of various exponents in the long-range polar ordered phase, as predicted by Toner \& Tu in 1995 (TT95) \cite{toner1995longrange,toner1998flocks}, measured by the largest-scale numerical calculations of the Vicsek model in 2019 \cite{mahault2019quantitative}, and measured in the Janus particle experiments \cite{iwasawa2021algebraic}. Since the upper critical dimension of the Toner-Tu equations is 4, the results coincide with the mean-field theory for $d\geq 4$. In the TT95 predictions, while the renormalization group calculations can be performed for $d=2$, for $d=3$ it is not possible to determine whether some terms are relevant, and thus there is no exact prediction of the exponents. The values shown here are the predicted values assuming that the same expressions are applicable for $d=3$ as for $d=2$. The numerical results of the Vicsek model and the experimental measurements of the Janus particles are directly measured without using theoretical scaling relationships. The exponent $\nu$ is difficult to measure even in numerical simulations. The numbers in parentheses indicate the uncertainties in the least significant digits.}
\begin{tabular}{r|ccc|cc|c}
\multicolumn{1}{l|}{} & \multicolumn{3}{c|}{$d=2$}                                                             & \multicolumn{2}{c|}{$d=3$}                       & $d\geq 4$ \\
\multicolumn{1}{l|}{} & \multicolumn{1}{c|}{TT95}   & \multicolumn{1}{c|}{Vicsek}             & Janus experiment          & \multicolumn{1}{c|}{TT95}    & Vicsek            & mean field       \\ \hline
real space correlation $\chi$            & \multicolumn{1}{c|}{$-1/5$} & \multicolumn{1}{c|}{$-0.31(2)$}         & $\simeq -0.26$ & \multicolumn{1}{c|}{$-3/5$}  & $\simeq -0.62$     & $1-d/2$   \\
anisotropy $\zeta$            & \multicolumn{1}{c|}{$3/5$}  & \multicolumn{1}{c|}{\textcolor[rgb]{0.7,0.3,0.1}{$0.95(2)$}}          & \textcolor[rgb]{0.7,0.3,0.1}{$\simeq 1$}      & \multicolumn{1}{c|}{$4/5$}   & \textcolor[rgb]{0.7,0.3,0.1}{$\simeq 1$}        & $1$       \\
wavenumber space correlation $\beta$           & \multicolumn{1}{c|}{$6/5$}  & \multicolumn{1}{c|}{$1.33(2)$}          &                & \multicolumn{1}{c|}{$8/5$}   & $1.77(3)$         & $2$       \\
decay of sound mode $z$               & \multicolumn{1}{c|}{$6/5$}  & \multicolumn{1}{c|}{$1.33(2)$}          & $\simeq 1.2$   & \multicolumn{1}{c|}{$8/5$}   & $\simeq 1.77$     & $2$       \\
GNF $\alpha$          & \multicolumn{1}{c|}{$4/5$}  & \multicolumn{1}{c|}{$0.84(1)$}          & $0.790(5)$      & \multicolumn{1}{c|}{$23/30$} & $0.79(2)$         & $1/2+1/d$   \\
transversal diffusion $\nu$               & \multicolumn{1}{c|}{$4/3$}  & \multicolumn{1}{c|}{$\simeq1.3$--$1.4$} &                & \multicolumn{1}{c|}{$1$}     & $\simeq 1$--$1.2$ & $1$      
\end{tabular}
\end{center}
}
\end{table*}

A quantitative comparison of the predictions by Toner-Tu 1995 \cite{toner1995longrange,toner1998flocks} and the results of numerical calculations of the Vicsek model with billions of particles \cite{mahault2019quantitative} is summarized in \tabref{tab:exponents}. As pointed out by Toner himself that the predicted values of the exponents may not be accurate \cite{toner2012reanalysis}, they do not match the numerical results. Notably, the anisotropy parameter $\zeta$ for the decays of the correlation functions is almost 1 in the numerical calculations, indicating very little anisotropy and a significant discrepancy with Toner-Tu 1995.

\subsection{Long-range order in Janus particles}
\label{sec:JanusLRO}
\begin{figure*}[tb]
\includegraphics[width=\textwidth]{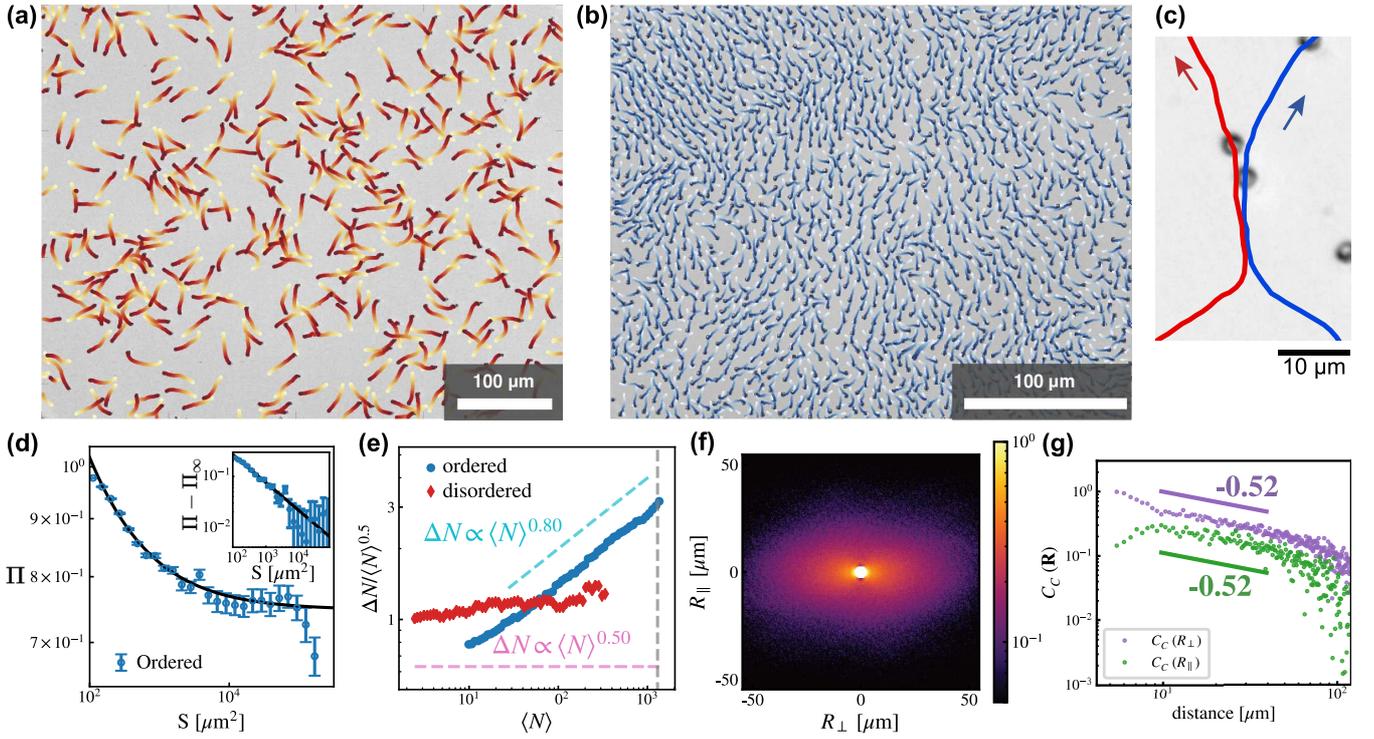}
\caption{Flocking transition in the electrokinetic Janus particles and statistical properties in their long-range ordered phase. In a high-frequency regime without adding NaCl to the solution with the condition corresponding to \figref{fig:JanusPhaseDiagram}(c), the Janus particles exhibit a flocking transition from (a) a disordered state with rotational symmetry at low density to (b) a long-range orientationally ordered state wit broken rotational symmetry. Trajectories of the particles are overlaid on the experimental snapshot with the colors changing from brighter to darker as the time elapses. The durations of the trajectories are 3 s and 2 s for the disordered and the ordered states respectively. (c) An example of trajectories of two interacting Janus particles showing that one particle is following the other particle due to electrostatic attraction. (d) Polar order parameter $\Pi$ as a function of the area $S$ of the region of interest in the ordered state. The best fit to an algebraic convergence to a positive finite value, signifying the existence of the true long-range order: $\Pi=\Pi_\infty+kS^{-\gamma/2}$ with $\Pi_\infty=0.74$, $k=3.1$, and $\gamma/2=0.53$. (e) Number fluctuations in the ordered state have a larger exponent $\alpha=-0.790(5)$ than in the disordered state, signifying the existence of GNF. A vertical gray dashed line indicates the scale of counterflows up to which the true long-range order is experimentally obtained.
(f) Color map of the orientation fluctuation correlation function $C_C(\bm{R})$ in the ordered state. The direction of the global order is set aligned with the vertical axis ($R_\perp$).
(g) Log-log plots of the slices of $C_C(\bm{R})$ both in the transverse ($R_\perp$, purple) and longitudinal ($R_\parallel$, green) directions. The solid lines are to guide the eyes. Isotropic decay with the anisotropy exponent $\alpha\simeq 1$ is observed.
Figures reproduced and modified from Ref.~\citen{iwasawa2021algebraic}.
}
\label{fig:JanusLRO}
\end{figure*}

As described in \secref{sec:Janus}, in a high-frequency low-ion-concentration regime, the Janus particles exhibit a flocking state, in which all the Janus particles move on average in the same direction \cite{iwasawa2021algebraic}(\figref{fig:JanusPhaseDiagram}(c)). This flocking state can be compared with the ordered state of the Vicsek model and the Toner-Tu theory.
It should be, however, noted that the correspondence between the experiments and the theoretical predictions needs to be assessed with care because the Vicsek model only considers short-range interactions while the electrokinetic Janus particles may exhibit longer-range hydrodynamic and/or electrostatic interactions. Indeed, some experimentally observed exponents differ from those the theoretical predictions \cite{toner1995longrange,toner1998flocks} and the numerical results \cite{mahault2019quantitative} of the Vicsek model, suggesting a different underlying mechanism responsible for the observed long-range order.

The Janus particles exhibit a flocking transition from a low-density, disordered state with rotational symmetry to a high-density, ordered state with broken rotational symmetry(\figref{fig:JanusLRO}(a),(b)). Such a flocking transition is realized through the competition between effective aligning interactions and fluctuations. With these experimental parameters, the electrostatic interactions between the two Janus particles attract their metal and dielectric hemispheres while relatively strong hydrodynamic interactions keep them apart, preventing chain formation. Consequently, one particle approaches and follows another particle during a collision event as shown in \figref{fig:JanusLRO}(c). This decreases the relative angle of the two particles' orientations, and this effective aligning interaction eventually gives rise to the global orientational order.

The presence of long-range orientational order in the flock of the Janus particles (\figref{fig:JanusLRO}(d)) is demonstrated through a finite-size scaling analysis of the polar order parameter $\Pi$, akin to the analysis performed in the Vicsek model (\figref{fig:VicsekModel_PhaseDiagram}(c)).
In this experiment, a suspension of the Janus particles was confined between the two ITO electrodes to form a large, pancake-shaped droplet with the diameter of a few centimeters.
The order parameter $\Pi$ was calculated by changing the area $S$ of the region of interest as in the case of the Vicsek model.
Although experimental observations focused at the center of the droplet to minimize boundary effects, counterflows of the particles inevitably occurred due to the particle number conservation within the closed droplet system, disrupting the polar orientational order.
Nonetheless, the polar order $\Pi$ exhibited a true long-range order behavior up to the scales of the counterflow, decaying more slowly than a power law. Therefore, this provides experimental evidence for the presence of the true long-range polar orientational order in 2D active matter systems.

The long-range ordered phase of Janus particles enabled the experimental extractions of several relevant exponents (see \tabref{tab:exponents}).
The presence of the GNF was reported with the exponent $\alpha=0.790(5)$ (\figref{fig:JanusLRO}(e)), close to the Toner-Tu 1995 prediction \cite{toner1995longrange,toner1998flocks} but smaller than that of the largest Vicsek model simulation \cite{mahault2019quantitative}.
Furthermore, the exponents corresponding to those in eqs.~\eqref{eq:TonerTuVelocityCorrelation}, \eqref{eq:TonerTuVelocityCorrelationFourier}, and \eqref{eq:TonerTuDensityCorrelation} were calculated. Because the polarity $\bm{n_p}$ of each Janus particle can be detected independently of its velocity, we can calculate an orientation fluctuation correlation function $C_C(\bm{R})$, which corresponds to $C_{v}(\bm{R})$ in the Vicsek model,
\begin{equation}
C_C(\bm{R}) \equiv \langle \langle \delta n_{p\perp}(t,\bm{r}) \delta n_{p\perp}(t,\bm{r}+\bm{R}) \rangle_{\bm{r}} \rangle_t,
\end{equation}
where $\delta n_{p\perp}(t,\bm{r}) = \bm{n_p}(t,\bm{r}) - \langle\bm{n_{p}\rangle}$ is the deviation of a single particle's polarity $\bm{n_p}$ from the global order $\langle\bm{n_{p}\rangle}$, and $\langle\cdot\rangle_{\bm{r}}$ is an average for all $|\bm{R}|$-distanced particle pairs. As shown in \figref{fig:JanusLRO}(f)(g), the correlation function decays algebraically in both directions with nealy identical exponents close to $-0.52$. This algebraic decay signifies the scale-free features of this colloidal flock that originate from the spontaneous symmetry breaking of the rotational symmetry. Importantly, the anisotropy of the decay is vanishingly small. As discussed in \ref{sec:VicsekScalingExponents}, this small anisotropy is consistent with the numerical results of the Vicsek model but deviates from the Toner-Tu predictions.
Comparing the results from the Janus particles with the scaling in the Toner-Tu theory, $C_C(\bm{R})\propto R_\perp^{2\chi}, R_\parallel^{2\chi/\zeta}$, gives the values of $\chi\simeq-0.26$ and $\zeta\simeq 1$ (\tabref{tab:exponents}). In addition to these equal-time correlation functions, the space-time correlations of the density fluctuations, defined in eq.~\eqref{eq:SpaceTimeDensityCorr}, captured sound modes propagating within the colloidal flocks. Their dispersion gave an estimate of the dynamical exponent $z\simeq 1.2$, which coincided with the Toner-Tu predictions.

Both quantitative and qualitative differences in the values of exponents, particularly the anisotropy $\zeta$, obtained from experimental, numerical, and theoretical results suggest the need for improvements in the theoretical framework.
For instance, it is possible that some nonlinearities neglected in the Toner-Tu calculations \cite{toner1995longrange,toner1998flocks} may be relevant asymptotically, as suggested in Ref.~\citen{mahault2019quantitative}. Together with the observation of different exponents in another experiment with flocking colloidal Quincke rollers \cite{geyer2018sounds}, these experimental and numerical results highlight the need for a quantitative understanding of how the details of the experimental systems affect the values of the exponents as well as to what extent we can expect universality.

Janus particles also exhibit intriguing properties beyond those observed in the Vicsek model and Toner-Tu predictions.
Unlike the Vicsek model, we can independently identify the polarity and velocity of each Janus particle in experiments \cite{nishiguchi2018flagellar,iwasawa2021algebraic}.
The polarity of each particle wihtin the flock was tracked, and its autocorrelation function was calculated to extract the rotational diffusion of the particle. The results indicated suppressed rotational diffusion in the ordered state for a long time scale, as expected, because the polarity of each particle is restored to the orientation of the global order. However, counterintuitively, it was found that, for a short time scale, the particles in the ordered state experience stronger rotational diffusion than those in the disordered state \cite{iwasawa2021algebraic}. This enhanced diffusion could be caused by frequent and continuous interactions among closely distanced particles in the ordered state. Moreover, the cross-correlation of the polarity and velocity of each particle revealed that the polarity follows the velocity \cite{iwasawa2021algebraic}, suggesting a rich interplay of these two properties that is absent in the Vicsek model.

\subsection{The Vicsek-style models with different symmetries}
\label{sec:VicsekStyleDifferentSymmetry}
\begin{figure}[tb]
\includegraphics[width=\columnwidth]{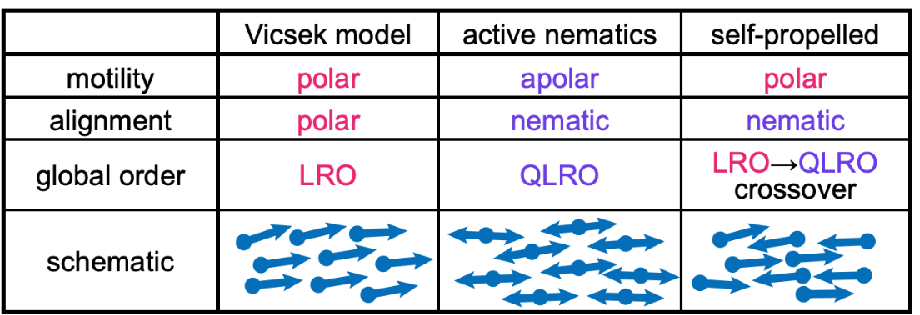}
\caption{Classification of the Vicsek-style models based on their symmetries. LRO and QLRO stand for true long-range order and quasi-long-range order, respectively.}
\label{fig:VicsekModel_classification}
\end{figure}

\subsubsection{Classification of the models}
The Vicsek model can be generalized to describe systems with different symmetries (\figref{fig:VicsekModel_classification}).
The original Vicsek model, which we have examined so far, deals with the collective motion of polar particles with polar interactions, meaning that each particle always moves in the direction of its own polarity and exhibits ferromagnetic alignment interactions. Actual experimental systems, however, include not only polar-polar systems such as the electrokinetic Janus particles but also systems with polar motility but apolar nematic interactions, such as swimming bacteria \cite{nishiguchi2017longrange} (\figref{fig:DimensionBacteria}). In addition, there exist systems like mammalian cells with elongated shapes \cite{kawaguchi2017topological} and vibrated rod-shaped particles \cite{narayan2007longlived} that align nematically and exhibit apolar motion due to stochastic reversal of their direction of motion. To describe such systems with different symmetries, several Vicsek-style models have been investigated.

In this context, ``Vicsek-style'' refers to a model that deals with point-like particles with short-range interactions as in the original Vicsek model.
The Vicsek-style models with the aforementioned symmetries are called self-propelled rods (polar motility with nematic interactions) and active nematics (apolar motility with nematic interactions) (\figref{fig:VicsekModel_classification}). 

\subsubsection{Controversy over long-range order in Vicsek-style models with nematic interactions}
\label{sec:Self-PropelledRods}

The two Vicsek-style models with nematic interactions -- active nematics and self-propelled rods -- exhibit flocking transitions from disordered states to some sort of long-range ordered states, similar to the original Vicsek model. These transitions are discontinuous, and coexistence phases have been observed \cite{chate2020dry}. These ordered phases commonly exhibit GNF, with exponents of $\alpha \simeq 0.8$.
However, performing renormalization group analysis similar to that in the original Vicsek model is far more challenging \cite{mishra2010dynamic}.
Hydrodynamic descriptions of active nematics and self-propelled rods contain the $\bf{Q}$-tensor, used in liquid crystal theories to describe the local nematic order, and thus their analytical treatment becomes much more complicated than that for the Toner-Tu equations. Therefore, theoretical predictions for exponents, such as the GNF exponent $\alpha$, for active nematics and self-propelled rods are, to the best of the Author's knowledge, obtained only numerically \cite{ginelli2010largescale,ngo2014largescale,mahault2021longrange} or from linear approximation theories \cite{narayan2007longlived, mahault2021longrange}.

One crucial difference lies in the types of long-range order observed in their ordered phases.
As we have already discussed, the Vicsek model exhibits true long-range order, where orientational order persists up to the infinite system size. In contrast, active nematics achieve only quasi-long-range order, in which the nematic order parameter decays to zero with a power-law dependence on the system size. This quasi-long-range order accompanies topological defects of the orientation field, akin to the equilibrium 2D XY model obeying the Hohenberg-Mermin-Wagner theorem.

The order in self-propelled rods has been controversial until recently. Initially, numerical simulations suggested true long-range order \cite{ginelli2010largescale}, and indeed the presence of true long-range order was experimentally demonstrated by using swimming bacteria in quasi-two dimensions (\figref{fig:DimensionBacteria}(a)), which will be described in more detail in \secref{sec:EcoliLRO}.
However, it has sometimes been argued that these systems should only exhibit quasi-long-range order, similar to active nematics, based on symmetry arguments.
Recent numerical simulations and hydrodynamic description analyses \cite{mahault2021longrange} have reported that self-propelled rods exhibit true long-range order but only up to an extremely large length scale that is usually inaccessible to experiments and even numerical simulations. Beyond this scale, the system crosses over to quasi-long-range order. This length scale is defined by a characteristic distance traveled between the direction reversals of the particles.

In the ordered state of the self-propelled rods model, the flock can be separated into two subgroups moving either to the left or right, assuming the global nematic order in the horizontal direction. The system of self-propelled rods was interpreted as a superposition of two interacting Vicsek models.
Due to the polar motility of self-propelled rods, the particles do not change their directions of motion and thus tend to stay in their own subgroups. Nevertheless, there are still possibilities that noise or interactions may induce flipping of the direction of motion and hopping between the two subgroups. This length scale defines the emergent macroscopic long-range order in the self-propelled rods, and practically we may only observe true long-range order behavior in experiments.

The decomposition into the two subgroups simplified the theoretical descriptions, allowing the construction of a hydrodynamic description for self-propelled rods \cite{mahault2021longrange}. As a result, analytical predictions for correlation functions and sound modes were obtained and compared with numerical results, giving numerical estimates of the exponents   (\tabref{tab:exponentsNematic}) \cite{mahault2021longrange}.  
A remarkable difference from the Toner-Tu theory is the value of the anisotropy parameter $\zeta$.
For the nematic flock of self-propelled rods in the true long-range order regime, $\zeta$ is reported to be $\zeta\simeq 1.25>1$, in stark contrast to $\zeta \lesssim 1$ reported for polar flocks in the numerical simulations of the original Vicsek model \cite{mahault2019quantitative}, the Toner-Tu theory \cite{toner1995longrange,toner1998flocks}, and the Janus experiment \cite{iwasawa2021algebraic}.
This opposite anisotropy is consistent with an experimental measurement done in the swimming bacteria in quasi-two dimensions \cite{NishiguchiSpringerTheses}, which will be discussed in \secref{sec:EcoliLRO}.

\begin{table*}[tb]
{
 \begin{center}
 \caption{\label{tab:exponentsNematic}
The values of exponents numerically measured in the true long-range nematic ordered phase (true LRO) and the quasi-long-range nematic ordered phase (quasi LRO) in $d=2$, and those measured experimentally in the bacterial true long-range nematic ordered phase. The numerical results for $\chi$, $\zeta$, $\beta$, and $z$ are from Ref.~\citen{mahault2021longrange}, where the crossover from true LRO to quasi LRO is reported. The numerical values of $\alpha$ are taken from Ref.~\citen{ginelli2010largescale} for true LRO and Ref.~\citen{ngo2014largescale} for quasi LRO. Experimental values of these exponents from bacterial experiments are based on Refs.~\citen{nishiguchi2017longrange} and \citen{NishiguchiSpringerTheses}.
}
\begin{tabular}{r|ccc}
\multicolumn{1}{l|}{} & \multicolumn{3}{c}{nematic phases in $d=2$} \\
\multicolumn{1}{l|}{} & \multicolumn{1}{c|}{true LRO (simulation)}   & \multicolumn{1}{c|}{quasi LRO (simulation)}             & \multicolumn{1}{c}{true LRO in 
bacteria (experiment)}          \\ \hline
real space correlation $\chi$            & \multicolumn{1}{c|}{$\simeq -0.25$} & \multicolumn{1}{c|}{$\simeq -0.45$}         & $-0.202(7)$  \\
anisotropy $\zeta$            & \multicolumn{1}{c|}{$\simeq 1.25$}  & \multicolumn{1}{c|}{$\simeq 1.1$}          & {$1.22(5)$}  \\
wavenumber space correlation $\beta$           & \multicolumn{1}{c|}{$\simeq 1.75$}  & \multicolumn{1}{c|}{$\simeq 1.2$}          & $1.9(1)$ $^{*1}$\\
decay of sound mode $z$               & \multicolumn{1}{c|}{$\simeq 1.75$}  & \multicolumn{1}{c|}{$\simeq 1.2$}          &  \\
GNF $\alpha$          & \multicolumn{1}{c|}{$\simeq 0.8$}  & \multicolumn{1}{c|}{$\simeq 0.8$}          & $\simeq 0.63$ $^{*2}$ \\
transversal diffusion $\nu$ $^{*3}$             & \multicolumn{1}{c|}{  }  & \multicolumn{1}{c|}{ } &   \\
\end{tabular}
\end{center}
\footnotesize{*1 While the exponents of $\hat{C}_v(\bm{q})$ and $\hat{C}_\rho(\bm{q})$ in Fourier space in the $\bm{q}_\perp$-direction ($|q_\||=0$) are both given by $-\beta$ in the Vicsek model, true LRO phase of self-propelled rods has different exponents. Specifically, the density correlation is modified to $\langle \delta \hat{\rho}(\bm{q},t)^2\rangle\sim q_\perp^{2-\beta}$ ( for $|q_\||=0$). Therefore, $\beta$ here is estimated from the orientation fluctuations of the bacterial director field, not from the density fluctuations.

*2 The exponent $\alpha$ in the bacterial experiment is obtained by binarizing the images, and thus may not be so precise as direct counting of bacteria. In fact, a Fourier space analysis of the equal-time density correlation functions leads to $\alpha\simeq0.8$, which is consistent with the numerical results \cite{NishiguchiSpringerTheses}.

*3 Due to the difficulty in estimating transversal diffusions, to the best of our knowledge, there has been no report on their exponent $\nu$ for the nematic cases.}
}
\end{table*}

\subsection{Long-range order in quasi-2D swimming bacteria}
\label{sec:EcoliLRO}
\begin{figure*}[tb]
\includegraphics[width=\textwidth]{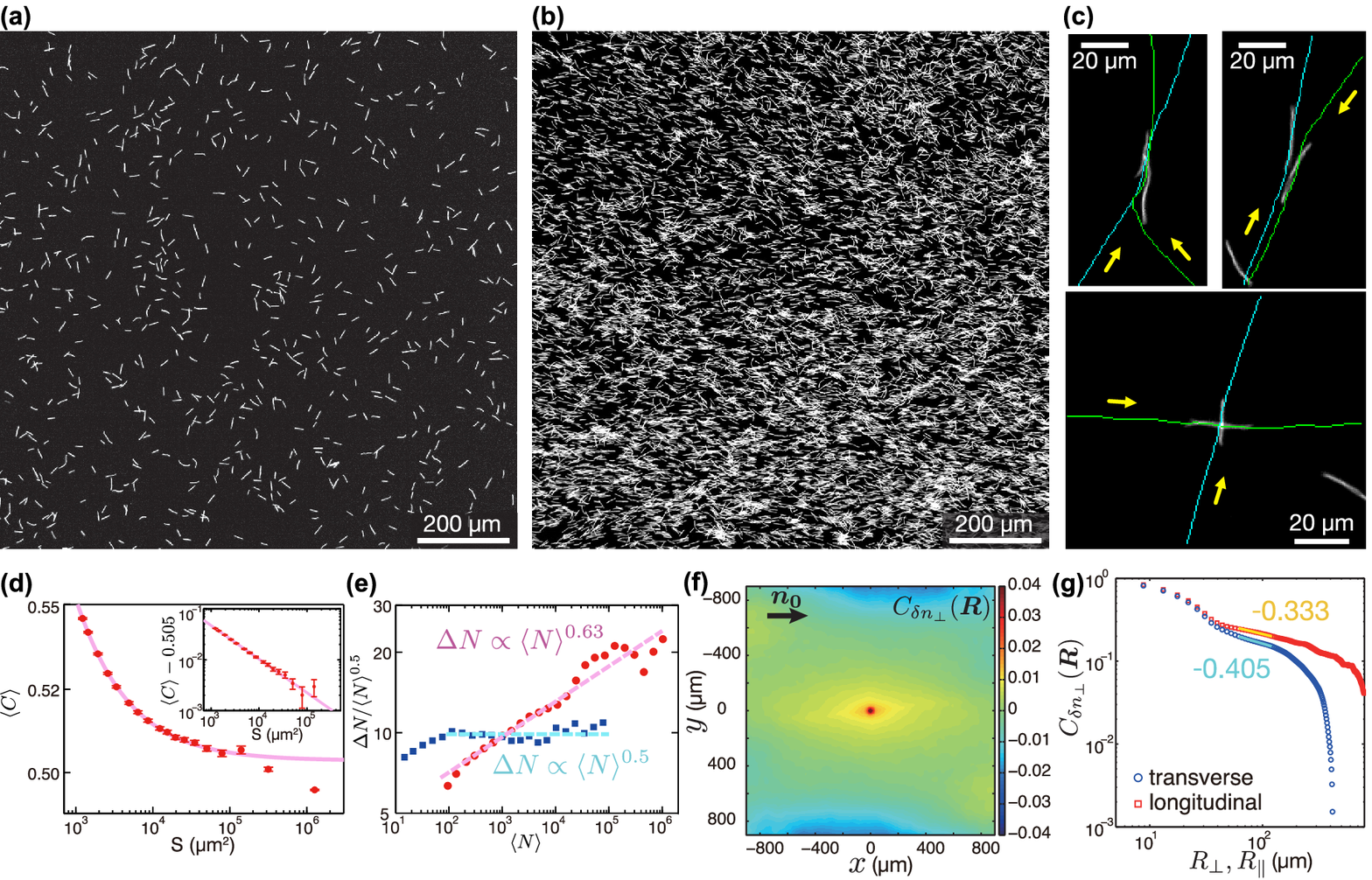}
\caption{Flocking transition in swimming filamentous bacteria confined in a quasi-2D fluid layer and the statistical properties in their long-range ordered phase.
(a) Disordered state at low density with rotational symmetry.
(b) Ordered state at high density with broken rotational symmetry.
(c) Trajectories of bacteria during collisions. Typical collisions induce nematic alignment. Bacteria with an acute incoming angle (top-left) result in parallel alignment, while those with an obtuse incoming angle (top-right) result in anti-parallel alignment. Due to the existence of tiny but finite $z$-dimention in the quasi-2D setup, bacteria can cross each other (top-left) and sometime hardly interact (bottom) during collisions.
(d) Average values of coherency $C$ corresponding to the nematic order parameter as a function of the area of the regions of interest. They show a decay slower than a power law with algebraic convergence to a positive value, signifying the existence of the true long-range order. The best fit gives $\langle C \rangle=C_\infty+kS^\gamma$ with $C_\infty=0.505$, $\gamma=-0.66$, and $k=4.6$.
(e) The ordered state exhibits GNF with an exponent $\alpha\simeq0.63$ significantly larger than in the disordered state. In the disordered state, normal fluctuations with $\alpha=0.5$ are confirmed. Note that number fluctuations here were estimated not by direct counting of the bacterial numbers but by the intensity fluctuations of the images.
(f) Colormap of the correlation function $C_{\delta n_\perp}(\bm{R})$. The direction of the global nematic order $\bm{n_0}$ is reorganized to the $x$-direction.
(g) Log-log plot of the slices of the correlation function $C_{\delta n_\perp}(\bm{R})$. Blue circles: the transverse direction, perpendicular to $\bm{n_0}$). Red squares:  the longitudinal direction, along $\bm{n_0}$. The estimated exponents are $-0.405(14)$ for the transverse direction (cyan solid line) and $-0.333(6)$ for the longitudinal direction (yellow solid line) respectively, signifying the anisotropy exponent $\zeta=1.22(5)$ greater than 1.
Figures (a), (b). (c), (d),(e), (f) reproduced and modified from Ref.~\citen{nishiguchi2017longrange}, and (g) from Ref.~\citen{NishiguchiSpringerTheses}
}
\label{fig:EcoliLRO}
\end{figure*}

As we have seen in \secref{sec:BacteriaDimension}, swimming bacteria in a thin quasi-2D fluid layer exhibit a nematically ordered state (\figref{fig:DimensionBacteria}(a),(c)) \cite{nishiguchi2017longrange}. Considering the symmetry, this system can be compared with the ordered state of the Vicsek-style self-propelled rods model as discussed in the previous section. For low enough density, bacteria swim in uncorrelated random directions (\figref{fig:EcoliLRO}(a)) due to the intrinsic isotropy of the system. At high density, the system exhibits a flocking state with global nematic order, in which bacteria swim in either direction of the nematic order (\figref{fig:EcoliLRO}(b)). Such an emergent nematic order is made possible due to the existence of the small third dimension, which is absent in strictly 2D systems. The gap width comparable to bacterial diameters in this quasi-2D system realizes interactions closer to the Vicsek-style models than in 2D systems. In quasi-2D, bacteria are allowed to cross each other during collisions, similarly to the point-like particles in the Vicsek-style models (\figref{fig:EcoliLRO}(c)), which is impossible in strict 2D systems \cite{zhang2010collective}. In addition, bacteria can still pass by each other with very little interaction (\figref{fig:EcoliLRO}(c)). This makes the system even closer to Vicsek-style models, in which such nonaligning events may occur due to the presence of noise.

This nematic state turned out to exhibit true long-range order by applying a finite-size scaling analysis, and GNF was also found therein \cite{nishiguchi2017longrange}. In this experiment, the quasi-two-dimensionality of the system unavoidably results in overlapping bacteria in the microscopy images, making it challenging to detect each bacterium and its orientation.
To overcome this difficulty, the coarse-grained nematic field was evaluated by using the structure tensor method \cite{rezakhaniha2012experimental}, which analyzes the steepest and least steep directions of the intensity gradients of images, thereby identifying the orientations of the elongated objects. This method was confirmed to give the estimates of not only the director field but also the scalar nematic order parameter of the nematic field \cite{nishiguchi2017longrange}. The quantity corresponding to the scalar nematic order parameter, $| \langle \exp{(2i\theta_j^t)\rangle_j |}$, is called coherency $C$ in this method, and the average values of coherency in the ordered state of this bacterial experiment as a function of the observation area size is plotted in \figref{fig:EcoliLRO}(d). This signifies that the nematic order decays slower than a power law, meaning the existence of true long-range order. This is consistent with the numerical prediction made prior to this experiment \cite{ginelli2010largescale} and the recent numerical and theoretical results demonstrating the crossover from true long-range order to quasi-long-range order at experimentally inaccessible large length scales \cite{mahault2021longrange}.
As for GNF, since detecting and counting each bacterium was challenging, number fluctuations were evaluated by the intensity of fluorescent bacteria. This analysis established that number fluctuations show significantly larger exponents in the ordered state than in the disordered state, demonstrating the presence of GNF (\figref{fig:EcoliLRO}(e)). Perhaps because of the indirect estimation of the number density, the experimentally obtained exponent $\alpha\simeq 0.63$ is somewhat smaller than the numerical predictions $\alpha\simeq0.8$  (\tabref{tab:exponentsNematic})\cite{ginelli2010largescale,chate2020dry, chate2022dry}. However, it is reported that a detailed analysis of the equal-time density correlation functions in Fourier space suggests $\alpha\simeq0.8$ by assuming the relation $\alpha=1/2+\beta/2d$ from the Toner-Tu prediction based on eq.~\eqref{eq:TonerTuDensityCorrelation} holds even for this experiment \cite{NishiguchiSpringerTheses}.

Correlation functions in the ordered state exhibited algebraic decays in either direction relative to the global nematic order but with a peculiar anisotropy.
Similarliy to the correlation functions $C_v(\bm{R})$ for the Toner-Tu theory \cite{toner1995longrange,toner1998flocks} and $C_C(\bm{R})$ for the Janus experiment \cite{iwasawa2021algebraic}, the two-point correlation function of director fluctuations  $\bm{\delta n_\perp}=\bm{n}-\bm{n_0}$ was calculated, with $\bm{n}$ the director field and $\bm{n_0}$ the direction of the global order, as,
\begin{equation}
C_{\delta n_\perp}(\bm{R}) := \langle \langle \delta n_\perp(\bm{r},t) \delta n_\perp(\bm{r}+\bm{R},t) \rangle_{\bm{r}} \rangle_t ,
\end{equation}
where $\delta n_\perp$ is a signed norm of $\bm{\delta n_\perp}$.
As shown in \figref{fig:EcoliLRO}(f)(g), it exhibits power-law behavior both parallel and perpendicular to $\bm{n}_0$.
These algebraic decays exemplify the presence of Nambu-Goldstone modes and therefore support that this ordered state arises as a result of the spontaneous breaking of the rotational symmetry of the system. 
Although the ranges of the algebraic behavior are limited due to experimental constraints, such as $\langle \delta n_\perp(t, \bm{r}) \rangle_{\bm{r}}=0$ making $C_{\delta n_\perp}(\bm{R})$ negative at large distances and artifact arising from the bacterial body shapes below the characteristic body length $\sim 20 \;\mathrm{\mu m}$, their power-law exponents were evaluated to be \cite{NishiguchiSpringerTheses},
\begin{equation}
C_{\delta n_\perp}(\bm{R}) \propto \left\{ \begin{array}{ll} R_\perp^{-0.405(14)}  \\ R_\parallel^{-0.333(6)}, \end{array} \right.
\label{eq:EcoliCorreleationExponents}
\end{equation}
where the digits inside the brackets mean uncertainty of 95~\% confidence level. This dependence gives the experimental estimate of $\chi=-0.202(7)$ and $\zeta=1.22(5)$. The obtained value of $\chi$
somehow coincides with the Toner-Tu prediction $\chi=-0.2$ for polar particles with polar interactions \cite{toner1995longrange,toner1998flocks} (\tabref{tab:exponents}) and is different but close to the recent numerical result $\chi\simeq-0.25$ for the true long-range order regime of self-propelled rods \cite{mahault2021longrange}. Importantly, the anisotropy parameter $\zeta=1.22(5)$ from this experiment is greater than one, which contrasts starkly with the results for polar particles with polar interactions: the weak anisotropy $\zeta\simeq 1$ for the numerical results of the Vicsek model and $\zeta= 0.6$ for the Toner-Tu theory (\tabref{tab:exponents}). This opposite anisotropy had seemed odd until the recent numerical and theoretical work \cite{mahault2021longrange} predicted $\zeta\simeq1.25$ for self-propelled rods (polar particles with nematic interactions) as discussed in \secref{sec:Self-PropelledRods}. Therefore, currently, the positive anisotropy $\zeta>1$ in the long-range ordered phase of polar particles with nematic interactions is supported both numerically and experimentally.

Quasi-two-dimensionality of the system, i.e. allowing the self-propelled elements to cross during collisions, is crucial for the emergence of long-range order in nematic systems.
This was pointed out in this experiment \cite{nishiguchi2017longrange,NishiguchiSpringerTheses} and has been supported by recent experiments with variable gap widths of a Hele-Shaw chamber \cite{ghosh2022cross}. Later studies on gliding assay experiments of microtubules and kinesins showing nematic interactions \cite{tanida2020gliding} and on numerical simulations of self-propelled rods \cite{shi2018selfpropelled} have also demonstrated that allowing the particles to cross leads to the emergence of true long-range orientational order in otherwise clustering or turbulent populations. Therefore, this quasi-2D property to allow crossing events can be regarded as a necessary condition for the emergence of true long-range order in systems of polar particles with nematic interactions. Note that this may not be the case for polar particles with polar interactions as we have observed the emergent long-range order in the 2D Janus particle system.

This experiment is regarded as the first experimental evidence for the existence of true long-range order with GNF in active matter systems, which exemplifies the scare-free fluctuating ordered phases of collective motion, i.e. the Toner-Tu-Ramaswamy phases.
As we have discussed in the Janus experiment (\secref{sec:Janus}), polar orientational order can be easily disrupted in closed systems. Therefore, implementing artificial periodic boundary conditions by fabricating racetrack-like microfluidic channels is desirable for polar flocks, as originally proposed by Toner \& Tu \cite{toner1998flocks}. Indeed, experiments with the colloidal Quincke rollers have been performed in such geometries, and a flocking transition was observed. While their early experiment had difficulty in detecting GNF \cite{bricard2013emergence}, later analyses successfully revealed the existence of GNF \cite{geyer2018sounds}.
For Janus particles, however, microfluidic structures distort the electric field, making it difficult to stably observe particle motion, and thus such periodic boundary conditions have not been realized so far. Nonetheless, by preparing a sufficiently large experimental system, they have successfully demonstrated the existence of long-range order in a range smaller than where counterflows can occur\cite{iwasawa2021algebraic}. In the case of quasi-2D swimming bacteria, which has the symmetry of self-propelled rods, providing reservoirs for bacterial suspensions around the observation area enabled bacteria to reorient and come back to the observation area. This facilitated the realization of the Toner-Tu-Ramaswamy phase, as it solved the destabilization of the order by counterflows arising from the particle number conservation. Therefore, long-range nematic order with GNF was realized earlier than long-range polar order.

\section{Conclusions and perspectives}
\label{sec:Conclusion}
We have reviewed the theoretical backgrounds for the emergent long-range orientatinal order in active matter systems as well as experimental investigations using microswimmer systems. Their comparisons have deepened the understanding of how orientational order emerges in active matter and given the directions of future theoretical developments. However, as outlined in \secref{sec:Introduction} and \secref{sec:Microswimmers}, active matter presents much richer phenomena that are not restricted to long-range orientational order. For further theoretical understanding of active matter systems, microswimmer systems will continue to play an important role.
For example, the electrokinetic Janus particles are particularly a promising system. They can easily be tuned to show different macroscopic behaviors. Furthermore, their polarity and velocity vectors can be defined independently as in the ABP but unlike in the Vicsek model with the degenerated polarity and velocity. Therefore, they may serve as a powerful tool for bridging different descriptions of active matter systems such as the ABP and the Vicsek-style models \cite{iwasawa2021algebraic}.

One of the future experimental challenges is to observe the coexistence of the ordered and disordered phases as seen in the Vicsek-style models. Although a propagating polar flock within a disordered phase has been observed in the colloidal roller experiment \cite{bricard2013emergence}, coexisting phases have not yet been reported in systems corresponding to the symmetries of active nematics or self-propelled rods. 
While the phase coexistence is characterized by propagating Vicsek wave in the original Vicsek model, that in the Vicsek-style models with nematic interactions exhibits instability called nematic chaos, where we observe continuous buckling, twisting, breaking up, and reconnection of bands of aligned particles parallel to the nematic order \cite{ginelli2010largescale,ngo2014largescale,chate2020dry,chate2022dry}.
Achieving experiments near the transition points to observe such spectacular coexistence requires precise control of the number density of the particles, which is challenging in most experiments. One promising method to realize such an experiment is to automatically sweep the density by proliferating bacterial population under perfusion \cite{shimaya2021scale, lama2022emergence}, which may allow experimental observation right on the transition points.

Although some experimental examples of emergent long-range order have been accumulated, a systematic understanding of the conditions under which long-range order emerges needs to be established.
As discussed in \secref{sec:EcoliLRO}, weak steric interactions, realized in quasi-2D environment that allows particles to cross over each other, seem to be key to achieving long-range nematic order, at least for otherwise turbulent or clustering 2D nematic systems such as swimming bacteria \cite{nishiguchi2017longrange} and gliding microtubules \cite{tanida2020gliding}. On the other hand, this does not seem necessary for long-range polar order because the particles cannot overlap each other in the Janus experiments \cite{iwasawa2021algebraic} as well as granular experiments numerically proven to exhibit long-range order under ideal conditions \cite{deseigne2010collective, weber2013longrange, kumar2014flocking, soni2020phases}. These examples suggest that long-range order in systems with polar interactions may be triggered via the following motion as in the Janus particles (\figref{fig:JanusLRO}(c)) and/or torque that aligns the polarity to the velocity of each particle as discussed in a vibrated polar disk systems \cite{lam2015selfpropelled}.
We note that similar contact following behavior has been reported in polar collective motion of mutant {\it Dictyostelium discoideum} population \cite{hayakawa2020polar}.

In this regard, to elucidate what kinds of interactions can give rise to long-range order, it may be helpful to infer the interactions in experiments by some machine learning or information-theoretic techniques \cite{frishman2020learning,zhang2021activeJanus} and relate them with the emergent order. However, inferring only binary interactions may not be sufficient to understand emergent order because it has been pointed out first in an actomyosin motility assay experiment \cite{suzuki2015polar} and later in the quasi-2D swimming bacteria \cite{NishiguchiSpringerTheses} that experimentally obtained binary-collision statistics cannot account for the growth of order parameters, i.e. the instability of disordered states, through the Boltzmann equation approach. As a matter of fact, in the ordered state of quasi-2D swimming bacteria, each bacterium is almost always in contact with at least one other bacterium, suggesting the importance of continuous and/or multi-particle collisions.

Although we have focused on long-range orientational order so far, there are other types of emergent order in collective motion. Therefore, it is also interesting to investigate how long-range order gets destabilized and then transits to another collective phase. For example, while the collective motion of quasi-2D swimming bacteria was interpreted in the context of the flocking transition between the dilute disordered state and the dense long-range ordered state in the Vicsek-style self-propelled rods model, it should also exhibit a route from the long-range ordered state to active turbulence by e.g. increasing the gap width with keeping the density constant. Since the Vicsek-style models do not describe active turbulence, investigating such transitions other than flocking transitions may pave the way for a better understanding of emergent order in active matter systems.
In relation to active turbulence, theoretical/numerical studies have revealed that different symmetries of the systems lead to different instabilities, resulting in different onset behaviors of turbulence when varying the activity of the systems \cite{giomi2011excitable,giomi2012polar,alert2022active}. 
In addition, active matter systems that develop active turbulence in the bulk \cite{dombrowski2004selfconcentration, blanch-mercader2017turbulent, sanchez2012spontaneous, alert2022active} can further be rectified to form stable vortices or directed flows under confinement or in the presence of obstacles \cite{wioland2013confinement, wioland2016directed, wioland2016ferromagnetic,beppu2017geometrydriven, nishiguchi2018engineering,reinken2020organizing,beppu2021edge, doxzen2013guidance, wu2017transition}. Detailed routes from ordered vortices to turbulence in these confined systems have recently been investigated by varying geometrical confinement of the system \cite{doostmohammadi2017onset,shiratani2023route}. Therefore, it may be interesting to see the connections among long-range ordered phases, active turbulence, and vortical orders by continuously changing the geometries, such as spatial dimensions, of bacterial systems, or by modifying the AC frequency or ion concentrations in the Janus experiment.

Exploring the effect of the spatial dimension on collective motion is of crucial importance also from the viewpoint of hydrodynamics.
A new mechanism has been proposed wherein an active hydrodynamic force may stabilize nematic order, as seen in the quasi-2D bacterial experiment \cite{nishiguchi2017longrange}, within a quasi-2D geometry \cite{maitra2018nonequilibrium}.
Another recent experimental/numerical study reported an anomalous hydrodynamic interaction in quasi-2D \cite{takaha2023quasitwodimensional}, revealing a sudden increase of hydrodynamic interaction between a single swimmer and obstacles when transitioning from a 3D to a quasi-2D environment.
Therefore, the understanding of such anomalous quasi-2D hydrodynamics may offer valuable insights into the impact of the spatial dimension on microswimmers' collective motion. To this end, defining a dimensionless parameter that quantifies spatial dimensions, applicable to different experimental setups, would help construct a universal understanding of collective motion.

Active matter experiments have posed important questions in statistical physics such as how and when long-range order is achieved in nonequilibrium systems, which has drawn attention from broad communities beyond the field of active matter physics.
Experimental explorations in active matter systems will continue to provide valuable insights into the fundamental principles of statistical physics, serving as a rich source of inspiration for future studies in physics and other disciplines such as materials science and biology.


\begin{acknowledgment}
The author thanks Beno\^it Mahault and Hugues Chaté for discussions while writing this review and for providing data, Alexandre Solon for providing the figures of the Vicsek model.
The author is also grateful to Igor S. Aranson, Markus B\"{a}r, Olivier B\'{e}nichou, Hugues Chat\'{e}, Vincent D\'{e}mery, Junichiro Iwasawa, Hong-Ren Jiang, Sabine H. L. Klapp, Beno\^it Mahault, Ken H. Nagai, Alexis Poncet, Henning Reinken, Andrey Sokolov, Masaki Sano, Alexey Snezhko, Sora Shiratani, Yuki Takaha, and Kazumasa A. Takeuchi for fruitful collaborations and discussions.
This work is supported in part by JSPS KAKENHI Grant Numbers 
JP19H05800, 
JP20K14426, 
and JP23H01141, 
and JST PRESTO ``Complex Flow'' Grant Number JPMJPR21O8, Japan.

\end{acknowledgment}

\profile{Daiki Nishiguchi}{was born in Osaka, Japan, in 1989, and grew up in Aichi, Saitama, Nagasaki, and finally Nara from the age of eight. In 2017, he earned his Ph.D. in physics from The University of Tokyo, and then pursued postdoctoral research at CEA-Saclay and Pasteur Institute in Paris for two years. Since 2019, he is an assistant professor at Department of Physics at The University of Tokyo. He enjoys dynamical phenomena both in biological and artificial systems under a microscope, with a particular focus on uncovering their universal laws and their ways of life.}

\bibliographystyle{jpsj}
\bibliography{Ref_JPSJreview}

\end{document}